%% file: root.tex
\documentclass[lettersize,journal]{IEEEtran}
\input{jared_macro}

\usepackage{array}
\usepackage{textcomp}
\usepackage{stfloats}
\usepackage{verbatim}
\def\BibTeX{{\rm B\kern-.05em{\sc i\kern-.025em b}\kern-.08em
    T\kern-.1667em\lower.7ex\hbox{E}\kern-.125emX}}
\usepackage{balance}
\begin{document}
\title{Bounding the Minimal Current Harmonic Distortion i\textit{}n Optimal Modulation of Single-Phase Power Converters}
\author{Jared Miller, Petros Karamanakos, \IEEEmembership{Senior Member, IEEE}, and Tobias Geyer, \IEEEmembership{Fellow, IEEE}
\thanks{J. Miller is with the Chair of Mathematical Systems Theory, Department of Mathematics,  University of Stuttgart, Stuttgart, Germany; (e-mail: jared.miller@imng.uni-stuttgart.de).}
\thanks{P. Karamanakos is with the Faculty of Information Technology and Communication Sciences, Tampere University, 33101 Tampere, Finland; (e-mail: p.karamanakos@ieee.org).}
\thanks{T. Geyer is with ABB System Drives, 5300 Turgi, Switzerland; (e-mail: t.geyer@ieee.org).}
\thanks{J. Miller  was partially supported by the Deutsche Forschungsgemeinschaft (DFG, German Research Foundation) under Germany's Excellence Strategy - EXC 2075 – 390740016. We acknowledge the support by the Stuttgart Center for Simulation Science (SimTech).}
}


\maketitle

\input{sections/abstract}

\begin{IEEEkeywords}
Optimal pulse patterns (OPPs), power converters, hybrid system, convex relaxations, harmonic distortions, pulse width modulation.
\end{IEEEkeywords}


\input{sections/acronym}
\input{sections/introduction}

\input{sections/motivation}

\input{sections/contributions}

\input{sections/summary}

\input{sections/preliminaries}

\input{sections/opp_as_ocp}
\input{sections/examples}
\input{sections/Conclusion}
\input{sections/acknowledgements}

\appendices
\input{appendix/app_energy}
\input{appendix/app_symmetry}
\input{appendix/app_measure_theory}
\input{appendix/app_convex_relaxation}

\bibliographystyle{IEEEtran}
\bibliography{references.bib}

\input{sections/biography}
\end{document}

%% file: jared_macro.tex
\usepackage[nolist]{acronym}
\usepackage{xcolor}
\usepackage{amsmath}
\usepackage{amssymb}
\usepackage{mathtools}
\usepackage{subcaption}
\usepackage{hyperref}
\hypersetup{colorlinks=true, linkcolor=black}
\usepackage[ruled]{algorithm2e}
\usepackage{mathrsfs}
\usepackage{circuitikz}
\definecolor{ccolor}{RGB}{203,96,21}

\newcommand{\urg}[1]{{\color{red} #1}}

\newcommand{\ct}[1]{\urg{[cite]}}
 
\newcommand{\R}{\mathbb{R}}

\newcommand{\N}{\mathbb{N}}

\newcommand{\Lie}{\mathcal{L}} 
 
\newcommand{\Z}{\mathbb{Z}}

\newcommand{\M}{\mathbb{M}}

\newcommand{\0}{\mathbf{0}}

\newcommand{\vs}{\mathcal{V}}
\newcommand{\es}{\mathcal{E}}
\newcommand{\gs}{\mathcal{G}}

\DeclarePairedDelimiter{\abs}{\lvert}{\rvert}
\DeclarePairedDelimiter{\norm}{\lVert}{\rVert}

\DeclarePairedDelimiterX{\inp}[2]{\langle}{\rangle}{#1, #2}

\DeclarePairedDelimiter{\Mp}{\mathcal{M}_+(}{)}

\DeclareMathOperator*{\argmax}{arg\!\,max}

\newtheorem{thm}{Theorem}[section]

\newtheorem{assum}[thm]{Assumption}

\newtheorem{prob}[thm]{Problem}

%% file: sections/abstract.tex
\begin{abstract}
\label{sec:abstract}%
%
Optimal pulse patterns (OPPs) are a modulation technique in which a switching signal is computed offline through an optimization process that accounts for selected performance criteria, such as current harmonic distortion. The optimization determines both the switching angles (i.e., switching times) and the pattern structure (i.e., the sequence of voltage levels). This optimization task is a challenging mixed-integer nonconvex problem, involving integer-valued voltage levels and trigonometric nonlinearities in both the objective and the constraints. We address this challenge by reinterpreting OPP design as a periodic mode-selecting optimal control problem of a hybrid system, where selecting angles and levels corresponds to choosing jump times in a transition graph. This time-domain formulation enables the direct use of convex-relaxation techniques from optimal control, producing a hierarchy of semidefinite programs that lower-bound the minimal achievable harmonic distortion and scale subquadratically with the number of converter levels and switching angles. Numerical results demonstrate the effectiveness of the proposed approach.
\end{abstract}

%% file: sections/acronym.tex
\begin{acronym}



\acro{FW}{Full-Wave}

\acro{GCT}{Gate-Commutated Thyristor}

\acro{HW}{Half-Wave}
\acro{LMI}{Linear Matrix Inequality}
\acroplural{LMI}[LMIs]{Linear Matrix Inequalities}
\acroindefinite{LMI}{an}{a}


\acro{LP}{Linear Program}
\acroindefinite{LP}{an}{a}
\acro{MPC}{Model Predictive Control}

\acro{OCP}{Optimal Control Problem}
\acroindefinite{OCP}{an}{a}

\acro{OPP}{Optimal Pulse Pattern}
\acroindefinite{OPP}{an}{a}



\acro{PSD}{Positive Semidefinite}

\acro{QW}[QaHW]{Quarter-and-Half-Wave}



\acro{SDP}{Semidefinite Program}
\acroindefinite{SDP}{an}{a}

\acro{SHE}{Selective Harmonics Elimination}
\acroindefinite{SHE}{an}{a}

\acro{SOS}{Sum of Squares}
\acroindefinite{SOS}{an}{a}

\acro{THD}{Total Harmonic Distortion}
\acro{TDD}{Total Demand Distortion}


\end{acronym}

%% file: sections/introduction.tex
\section{Introduction}
\label{sec:introduction}


\IEEEPARstart{P}{ower} converters are a fundamental technology in electrical energy conversion. They play a crucial role in a wide range of applications, including renewable energy integration and various industrial processes~\cite{zhong2012control, Jahns2001drives}. These converters consist of semiconductor devices that are actively controlled to transform electrical power from one form to another, e.g., from dc to ac. By appropriately modulating the switching instants, the ac output voltage can be synthesized as a waveform composed of a finite number of discrete voltage levels~\cite{kouro2010recent}.


However, the switching nature of power converters inherently introduces harmonic distortions, which degrade the quality of the output voltage and current. Therefore, the selection of the switching signal, i.e., the sequence of pulses applied to the converter, is crucial for minimizing these distortions~\cite{holtz2002pulsewidth}. Other performance metrics, such as power losses, common-mode voltage, and the amplitude of specific harmonics, must also be considered as they influence both the operation of the converter and its interaction with the grid or an electrical machine. Achieving an optimal trade-off among these criteria makes the choice of the pulse width modulation (PWM) strategy a key design factor~\cite{holmes2003pulse}.

Conventional PWM methods include carrier-based PWM (CB-PWM) and space vector modulation (SVM)~\cite{holmes2003pulse}. Although several variants of these techniques exist, their switching instants are deterministic. As a consequence, achieving the best possible balance among the aforementioned performance criteria can be challenging. In contrast, programmed PWM methods, such as selective harmonic elimination (SHE)~\cite{patel1973generalized,patel1974generalized} and optimal pulse patterns (OPPs)~\cite{buja1980optimum}, are computed offline, allowing specific design objectives to be explicitly incorporated into the computation procedure. As a result, their performance can be tailored to the requirements of a given application.


More specifically, the SHE-based switching signal is computed over one fundamental period by assuming specific symmetry properties, such as quarter- and half-wave (QaHW) symmetry. The switching instants (i.e., switching angles) are obtained by solving a set of nonlinear equations that enforce the desired harmonic characteristics~\cite{dahidah2014review}. For a fixed switching sequence structure (i.e., the sequence of pulses at specific voltage levels), these equations become polynomial relations in the sines and cosines of the unknown switching angles. Assuming $k$ switching angles to be determined, one equation ensures that the waveform synthesizes the desired fundamental component (i.e., the modulation index), while the remaining $k-1$ degrees of freedom are used to eliminate targeted harmonics. Since multiple solutions typically exist, one feasible set of angles is selected according to additional criteria, commonly the minimization of total demand distortion (TDD)~\cite{chiasson2004unified}. However, this sequential design procedure does not guarantee globally minimal distortion or full utilization of the available dc-link voltage. Moreover, as the number of unknown switching angles increases, the nonlinear problem may become infeasible or may yield infinitely many solutions when the constraints are degenerate~\cite{wells2005selective}.

OPPs, on the other hand, are computed through a mathematical optimization procedure. The objective function is typically chosen to represent the output current TDD, while additional performance goals can be incorporated either as extra terms in the objective function or as explicit constraints. Owing to this design flexibility, OPPs can achieve superior trade-offs between current distortions and other relevant metrics, such as limited common-mode voltage~\cite{koukoula2024optimal}, reduced power losses~\cite{geyer2023optimized,koukoula2025losses}, bounded junction temperature~\cite{dorfling2024jucntion,koukoula2025junction}, or compliance with grid harmonic standards~\cite{rahmanpour2023three}. Moreover, symmetry assumptions imposed on the switching signal can be easily relaxed to enlarge the feasible space and further improve performance~\cite{birth2019generalized}. 

Nevertheless, computing OPPs is highly nontrivial, as the underlying optimization problem is (mixed-integer) nonconvex. The integer part originates from the possible optimization of the pulse sequence (i.e., the sequence of applied voltage levels), while the nonconvexity stems from the trigonometric dependence on the unknown switching angles. Consequently, a wide variety of numerical methods has been explored for OPP computation, including iterative approaches~\cite{meili2006optimized}, genetic algorithms~\cite{dahidah2008hybrid}, swarm optimization~\cite{kavouski2012bee}, virtual-angle techniques~\cite{koukoula2024fast}, gradient-mode optimization~\cite{ali2024optimal}, neural-network-based approximation~\cite{toubal2022neural}, deep reinforcement learning~\cite{qashqai2020new,abu2025optimized}, and differentiable programming~\cite{abu2025diff}. However, none of these methods can guarantee that the obtained solution is globally optimal.

To provide a different perspective on the computation of OPPs, this paper adopts convex relaxation techniques from optimization and optimal control to derive lower bounds on the minimal current TDD produced by OPPs. The key idea is that the nonconvex OPP problem can be rendered convex by lifting it to an infinite-dimensional optimization problem in measures~\cite{rubio1975generalized, lewis1980relaxation, lasserre2009moments}, which must then be truncated into a sequence of finite-dimensional convex optimization problems to enable tractable computation. Existing truncation approaches include gridding methods~\cite{mohajerin2018infinite} and satisfiability modulo theory solvers based on interval propagation~\cite{gao2013dreal,abate2021fossil}. When all problem data can be expressed in polynomial form, the moment-sum of squares
(SOS) hierarchy~\cite{lasserre2009moments} provides a systematic alternative by approximating the lifted linear program through a sequence of semidefinite programs (SDPs) of increasing polynomial degree, whose optimal values yield a monotonically nondecreasing sequence of lower bounds to the optimum of the original nonconvex problem. The moment-SOS framework has been successfully applied to a variety of hybrid or switched-system problems, including barrier functions for safety~\cite{prajna2004safety}, optimal control~\cite{zhao2017optimal}, peak estimation~\cite{miller2023peakhy}, mode-selecting control~\cite{claeys2016modal}, risk estimation~\cite{miller2024chancepeak}, regions of attraction for periodic orbits~\cite{manchester2011regions}, and joint spectral radius approximation~\cite{parrilo2008approximation, ahmadi2019polynomial, wang2021sparsejsr}.

In the specific context of three-phase OPPs, the moment-SOS hierarchy has been applied in~\cite{wachter2021convex} to lower-bound the TDD for a given pulse pattern structure. In that work, both the objective function and the nonlinear harmonic constraints are first approximated using high-order Taylor series expansions. These are then represented by low-order minimax interpolating polynomials (e.g., degree 3–4) with valid error bounds. The resulting TDD bound converges to the true optimum as both the moment-SOS polynomial degree and the approximation degrees increase. The computational complexity of this approach grows polynomially with the approximation order and the number of switching angles.

%% file: sections/motivation.tex
Motivated by the above---and in contrast to typical frequency-domain methods---our preliminary work \cite{miller2025oppcdc} treated the OPP as a mode-selecting optimal control problem (OCP) of a hybrid system~\cite{teel2012hybrid}. The formulation is entirely in the time domain. In this way, choosing switching angles and switching levels (i.e., the pulse-pattern structure) becomes selecting the times and arcs to traverse in a transition graph. This hybrid-system perspective yields an infinite-dimensional linear program with polynomial data, which enables a direct application of the moment–SOS hierarchy without Taylor or interpolation approximations. The resulting scheme scales subquadratically with the product of the number of switching transitions and converter levels, and polynomially with the chosen approximation degree.

%% file: sections/contributions.tex
The present work builds on and expands \cite{miller2025oppcdc} by relaxing symmetry assumptions to introduce additional design freedom and improve the output current harmonic distortions. We also describe the convex relaxations in greater detail. Finally, the method is tested on a more realistic power electronic setup that can be interpreted as a grid-connected converter or a variable-speed drive, while the load resistance, which is typically omitted, is explicitly incorporated into the optimization. The contributions of this work are:
\begin{itemize}
    \item An interpretation of OPP as a mode-selecting hybrid-systems optimal control problem.
    \item The use of convex relaxation methods in optimal control to lower-bound the minimal TDD.
    \item Numerical results for bounding and controlling multi-level inverters.
\end{itemize}
To the best of our knowledge, the preliminary paper \cite{miller2025oppcdc} and this extended paper are the first to pose the OPP problem as an OCP and to apply OCP methods to bound the minimal TDD under constraints.

%% file: sections/summary.tex
The remainder of the paper is structured as follows. Section~\ref{sec:preliminaries} reviews the notation, the converter model, the OPP, and the associated optimization problem. Section~\ref{sec:opp_as_ocp} interprets the OPP as a periodic mode-selecting OCP in a hybrid system and explains how convexification yields lower bounds on the minimal current TDD. Numerical results that verify the effectiveness of the proposed method are presented in Section~\ref{sec:examples}. Finally, Section~\ref{sec:conclusion} concludes the paper and outlines future directions.

%% file: sections/preliminaries.tex
\section{Preliminaries}
\label{sec:preliminaries}

\subsection{Notation}
The $\ell$-dimensional space of real numbers is $\R^\ell$.
The set of integers is $\Z$, the set of nonnegative integers is $\Z_{\geq0}$, and the set of natural numbers is $\N$. The set of integers between $a$ and $b$ (inclusive) is $a..b$. Given two vectors $c_1, c_2 \in \R^\ell$, the partial order $c_1 \leq c_2$ will hold if $\forall i \in 1..\ell: c_{1i} \leq c_{2i}$. The same elementwise partial order is defined with respect to the comparators $=$ and $\geq$.

The map $\psi(\theta)$ performs a trigonometric lift with $\psi(\theta) = (\cos(\theta), \sin(\theta))$. The two-dimensional unit sphere is $B = \{(c, s) \in \R^2 \mid c^2+s^2 = 1\}$. Given angles $0 \leq \alpha_1 \leq \alpha_2 \leq 2 \pi$, the symbol $B(\alpha_1, \alpha_2)$ will denote the arc of the unit sphere between angles $\alpha_1$ and $\alpha_2$ ($B(\alpha_1, \alpha_2) = \{\psi(\theta) \mid \theta \in [\alpha_1, \alpha_2]\}$. 

The signal energy of a piecewise continuous $2\pi$-periodic  real-valued function $z(\theta)$ is $\norm{z}_2^2 = \int_{0}^{2\pi} z(\theta)^2 d \theta$.

\subsection{Model of the Power Electronic System}

Let $f_1$ denote the fundamental frequency, $\omega_1 = 2\pi f_1$ be the corresponding angular frequency, and $T_1 = 1/f_1$ be the fundamental period. In the sequel, all signals are assumed to be $2\pi$-periodic and are expressed with respect to an angle $\theta$, where $\theta = 0$ marks the beginning of the fundamental period and $\theta = 2\pi$ its end.

\begin{figure}[t!]
    \centering
    \begin{tikzpicture}
	\draw (9.25, 6) to[cute inductor, l={$L_{\text{load}}$}, label distance=0.02cm] (11, 6);
	\draw (11, 6) to[american resistor, l={$R_{\text{load}}$}, label distance=0.02cm] (13, 6);
	\draw (8, 4) to[square voltage source, l={$v_{\text{conv}}$}, label distance=0.02cm] (8, 6);
	\draw node[ground] at (11, 4) {};
	\draw (8, 6) -| (9.248, 6.001);
	\draw (8, 4) -- (9.25, 4);
	\draw (11, 4) -- (13.25, 4);
	\draw (9.25, 4) |- (11, 4);
	\draw (13, 6) -- (13.25, 6);
	\draw (13.25, 6) to[sinusoidal voltage source, l={$v_{\text{load}}$}, label distance=0.02cm] (13.25, 4);
	\draw[-stealth, line width=2pt] (9.25, 5.5) -- (11, 5.5) node[midway,below]{$i_{\text{load}}$};
\end{tikzpicture}
    \caption{Equivalent circuit of the considered power electronic system \label{fig:motor_current}}
\end{figure}
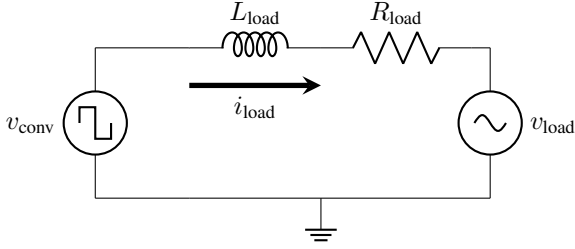

The considered power electronic system is shown in Fig.~\ref{fig:motor_current}. The converter output voltage is represented by the source $v_{\text{conv}}$, which supplies an active resistive–inductive load. This load may represent either the grid or an electrical machine, where $R_{\text{load}}$ denotes the resistance, $L_{\text{load}}$ the inductance, and $v_{\text{load}}$ the grid voltage or the back electromotive force (back-EMF) of a machine. A typical control objective is to generate a converter voltage such that the load current $i_{\text{load}}$ tracks a reference $i^*$ as closely as possible, thereby ensuring low ripple, i.e., low harmonic distortion. This is achieved by appropriately manipulating the converter switches.


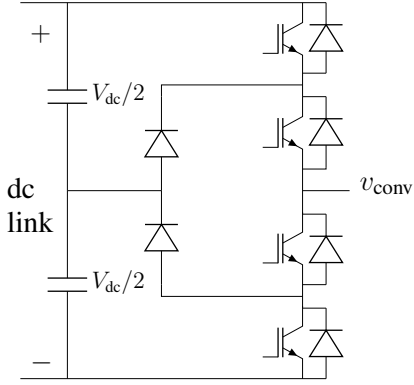
\begin{figure}[t!]
    \centering
    \resizebox{0.6\linewidth}{!}{%
\begin{tikzpicture}
	\draw (1, 9) to[capacitor, l={\Large{$V_{\text{dc}}/2$}}] (1, 5);
	\draw (1, 5) to[capacitor, l={\Large{$V_{\text{dc}}/2$}}] (1, 1);
	\draw (0, 9) -- (6, 9);
	\draw (0, 1) -- (6, 1);
	\draw (1, 5) -- (3, 5);
	\draw (3, 5) to[empty diode] (3, 7);
	\draw (3, 3) to[empty diode] (3, 5);
	\node[Lnigbt] at (6, 8.23){};
	\node[Lnigbt] at (6, 6.23){};
	\node[Lnigbt] at (6, 3.77){};
	\node[Lnigbt] at (6, 1.77){};
	\draw (6, 7.25) -- (3, 7.25) -| (3, 7);
	\draw (3, 3) -| (3, 2.75) -- (6, 2.75);
	\draw (6, 7.46) -| (6, 7);
	\draw (6, 5.46) -| (6, 4.54);
	\draw (6, 2.5) -| (6, 3);
	\draw (6, 5) -- (7, 5);
	\draw (6.5, 7.5) to[empty diode] (6.5, 9);
	\draw (6.5, 5.5) to[empty diode] (6.5, 7);
	\draw (6.5, 3) to[empty diode] (6.5, 4.5);
	\draw (6.5, 1) to[empty diode] (6.5, 2.5);
	\draw (6.5, 1) -- (6, 1);
	\draw (6.5, 2.5) |- (6, 2.5);
	\draw (6.5, 3) -- (6, 3);
	\draw (6.5, 4.5) -| (6, 4.54);
	\draw (6.5, 5.5) |- (6, 5.46);
	\draw (6.5, 7) -- (6, 7);
	\draw (6.5, 7.5) -- (6, 7.5);
	\draw (6.5, 9) -- (6, 9);
	\node[shape=rectangle, minimum width=0.965cm, minimum height=0.715cm] at (7.5, 5.125){} node[anchor=north west, align=left, text width=0.577cm, inner sep=6pt] at (7, 5.5){\Huge{$v_{\text{conv}}$}};
	\node[shape=rectangle, minimum width=0.715cm, minimum height=0.715cm] at (0.375, 8.375){} node[anchor=north west, align=left, text width=0.327cm, inner sep=6pt] at (0, 8.75){\Huge{$+$}};
	\node[shape=rectangle, minimum width=0.715cm, minimum height=0.715cm] at (0.375, 1.375){} node[anchor=north west, align=left, text width=0.327cm, inner sep=6pt] at (0, 1.75){\Huge{$-$}};
	\node[shape=rectangle, minimum width=0.965cm, minimum height=1.215cm] at (0, 4.875){} node[anchor=north west, align=left, text width=0.577cm, inner sep=6pt] at (-0.5, 5.5){\Huge{dc \\[8pt] link}};
\end{tikzpicture}%
}
    \caption{Single-phase voltage-source converter with three output voltage levels and four active switches}
    \label{fig:inverter_diagram}
\end{figure}

Specifically, the converter is assumed to generate $N$ discrete voltage levels, whose admissible values are collected in the set $L = \{u_n\}_{n=1}^N$. For a dc-link voltage $V_{\text{dc}}$, the converter output voltage therefore satisfies
$v_{\text{conv}}(\theta) \in \left\{ u_n \,\frac{V_{\text{dc}}}{2} \right\}_{n=1}^N$
As an example, Fig.~\ref{fig:inverter_diagram} illustrates a three-level converter, whose output voltage levels are $-V_{\text{dc}}/2$, $0$, and $V_{\text{dc}}/2$, corresponding to switch positions $u \in \{-1,0,1\}$, respectively.


Given this setup, the load current dynamics are described by
\begin{align}
    \frac{\mathrm{d}{i}_{\text{load}}(\theta)}{\mathrm{d}\theta}
    &= \frac{1}{L_{\text{load}}}\!\left(
        \frac{V_{\text{dc}}}{2}u(\theta)
        - v_{\text{load}}(\theta)
        - R_{\text{load}}\, i_{\text{load}}(\theta)
    \right).
    \label{eq:dynamics}
\end{align}
For convenience, we introduce normalized coordinates $i = (2L_\text{load}/V_{\text{dc}})\, i_{\text{load}}$, and $u_{\text{load}} = (2L_\text{load}/V_{\text{dc}})\, v_{\text{load}}$, which yield the simplified expression
\begin{align}
    \frac{\mathrm{d}{i}(\theta)}{\mathrm{d}\theta}
    = u(\theta) - \tilde{u}_{\text{load}}(\theta) - \tau\, i(\theta),
    \label{eq:dynamics_per_unit}
\end{align}
where $\tau = R_{\text{load}}/L_{\text{load}}$ is the load time constant, and $\tilde{u}_{\text{load}} = u_{\text{load}}/L_{\text{load}}$.

%

\subsection{Optimal Pulse Patterns}

OPPs are computed such that a chosen performance criterion, such as the load current TDD, is minimized while satisfying given constraints. In this context, an OPP is a $2\pi$-periodic switching signal $u(\theta)$ with $k$ switching transitions per fundamental period, where $k$ is always even due to the $2\pi$-periodicity. The pulse number $d$ associated with the pattern is defined as the half-integer $d = k/4$. In the case of a three-level converter, this implies an average device switching frequency of $f_{\text{sw}} = k f_1$.

An OPP can be described by the switching angles $\{\alpha^i\}_{i=1}^{k} \in [0, 2\pi]^k$ and the switch positions $\{u^i\}_{i=0}^{k} \in L^{k+1}$ through the piecewise constant function
\begin{align}
\label{eq:v_theta_step}
    u(\theta) &= \begin{cases}  u^0 & \theta \in [0, \alpha^1) \\      
        u^i & \theta \in [\alpha^{i}, \alpha^{i+1}), \ i \in 1..k-1\\        
        u^k & \theta \in [\alpha^k, 2\pi]
    \end{cases}.
\end{align}
This signal admits the Fourier series
\begin{align}
    u(\theta) &= \frac{a_0}{2} + \sum_{\ell=1}^\infty a_\ell \cos(\ell \theta ) + b_\ell \sin(\ell \theta) ,
\end{align}
with the Fourier coefficients \cite[Eq. (8)]{birth2019generalized}
\begin{subequations}
\label{eq:fourier_coeff}    
\begin{align}
    a_0 &= 2 {u}^0 - \frac{1}{\pi} \sum_{i=1}^k ({u}^{i} - {u}^{i-1}) {\alpha}^i& b_0 &= 0 \\
    a_\ell &= \frac{-1}{ \ell \pi} \sum_{i=1}^k ({u}^{i} - {u}^{i-1}) \sin(\ell {\alpha}^i) \\
    b_\ell & = \frac{1}{ \ell \pi} \sum_{i=1}^k ({u}^{i} - {u}^{i-1}) \cos(\ell {\alpha}^i).
\end{align}
\end{subequations}
These Fourier coefficients are nonlinear functions of $u$ and $\alpha$.

Assuming that the externally applied normalized signal is of the form
\begin{align}
    \tilde{u}_{\text{load}}(\theta) =  A \cos(\theta + \phi),
\end{align}
then the Fourier coefficients $(\tilde{a}, \tilde{b})$ of the load current $i_\text{load}$ are
\begin{subequations}
\label{eq:fourier_load}
\begin{align}
    \tilde{a}_1 + j \tilde{b}_1 &=  \frac{1}{R_{\text{load}} + j \omega_1 L_{\text{load}}}((a_1 + j b_1) + A \exp(j \phi)) \\
    \tilde{a}_{\ell } + j \tilde{b}_{\ell } &=  \frac{1}{(R_{\text{load}} + j \omega_1 L_{\text{load}})}(a_{\ell } + j b_{\ell}) \qquad \forall \ell \neq 1.
    \end{align}
\end{subequations}
After defining a constant of proportionality 
\begin{align}
    C_p = \frac{1}{\sqrt{2}I_{\text{nominal}} \omega_1 \sqrt{R_{\text{load}}^2 + L_{\text{load}}^2}  } \frac{V_{\text{dc}}}{2},
\end{align}
the load current TDD is given by
\begin{align}
    \text{TDD}[I_\text{load}] &= C_p\sqrt{\sum_{\ell \neq 1} \tilde{a}_\ell^2 + \tilde{b}_\ell^2}. \label{eq:tdd_fourier}
\end{align}
According to the above expression, the TDD is zero if all signal energy resides in the fundamental component, i.e., the signal is purely sinusoidal. The TDD is therefore a measure of spectral efficiency. 
By Parseval's relation, the formula in \eqref{eq:tdd_fourier} can be transformed into a time-domain expression
\begin{align}
     \text{TDD}[I_\text{load}] &=  C_p \sqrt{\norm{I}^2_2/\pi - \tilde{a}_1^2 - \tilde{b}_1^2} = C_p \text{TDD}[I].
\end{align}

Due to this $C_p$-proportionality, minimizing $\text{TDD}[I]$ will minimize $\text{TDD}[I_{\text{load}}]$. 
In the case where $\tilde{a}_1$ and $\tilde{b}_1$ are fixed, monotonicity of the square root implies that  minimizing $\norm{I}_2^2$ will correspondingly minimize $\text{TDD}[I]$ and $\text{TDD}[I_{\text{load}}]$.

The signal energy $\norm{I}_2^2$ has a closed-form expression in terms of the switching angles $\alpha$ and the initial current $i(0)$, assuming a fixed switching sequence $\{u^{i}\}$ and considering only the pulse-driven component of the current (i.e., neglecting the contribution of the load source). The explicit formula for $\norm{I}_2^2$ is provided in~\eqref{eq:energy_total_label} in Appendix~\ref{app:energy}. When $\tau = 0$ and $A = 0$, this expression reduces to a piecewise cubic polynomial in $\alpha$; otherwise, it is a piecewise nonlinear function of $\alpha$.

\subsection{Symmetries}

\begin{figure}[t!]
    \centering
    \includegraphics[width=0.7\linewidth]{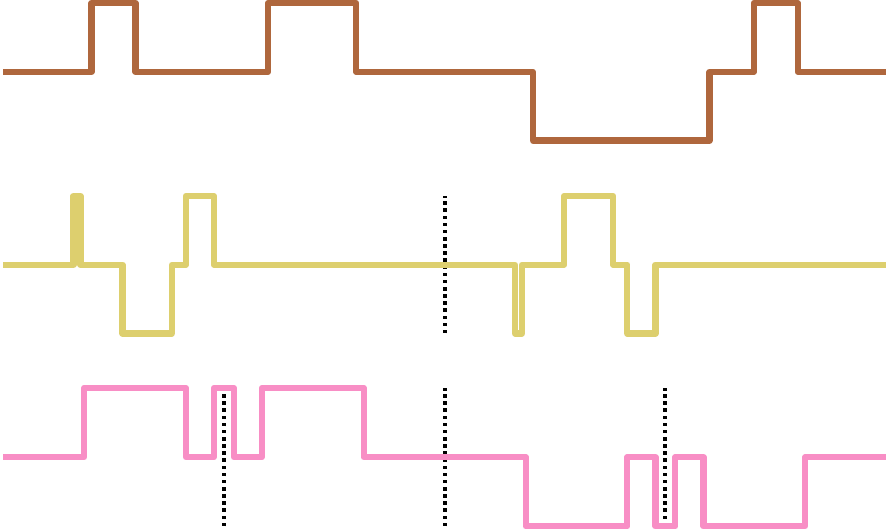}
    \caption{Example of patterns with (from top to bottom) \ac{FW}, \ac{HW}, and \ac{QW} symmetries}
    \label{fig:symmetries}
\end{figure}

Symmetries can be imposed on the switching signal $u$ to constrain harmonics and reduce the complexity of the associated OPP problem. The three main types of symmetries are (see Fig.~\ref{fig:symmetries})
%
\begin{subequations}
\label{eq:symmetries}
\begin{align}
& \text{Full-wave} &  u(\theta+2\pi) &=u(\theta) & & \forall \theta \in \R \\
    & \text{Half-wave} &  u(\theta+\pi) &= -u(\theta) & & \forall \theta \in [0, \pi] \\    
    & \text{Quarter-wave} &  u(\theta) &= u(\pi - \theta) & & \forall \theta \in [0, \pi].
\end{align}
\end{subequations}
Both half-wave (HW) and quarter-wave (QaHW) symmetry imply that $u(\theta)$ has zero mean, i.e., its dc component is zero. Moreover, HW symmetry forces all even-order harmonics to vanish, while QaHW symmetry imposes that all cosine coefficients satisfy $a_\ell = 0$ for every $\ell$. Finally, a pulse pattern with HW (respectively QaHW) symmetry is uniquely determined by the first $k/2$ (respectively $k/4$) switching angles and their associated transitions. 

\subsection{Constraints in Optimal Pulse Patterns}

\begin{table}[t!]
    \centering
        \caption{Constraints in OPP Problem}
    \begin{tabular}{cl}         
         Design & Constraint Description \\ \hline                     
           \checkmark & Number of switching angles \\
           \checkmark & Symmetry \\
           \checkmark & Unipolarity \\
          & Harmonics specification  \\
           & Interlocking angle \\
           & Converter levels \\
           & Adjacency restriction \\
           & Periodicity 
    \end{tabular}
    \label{tab:constraints}
\end{table}

\begin{table}[t!]
    \centering
    \caption{Design Choices}
    \begin{tabular}{c l}
    $k$ & Number of switching angles \\
         Sym & Symmetry (\ac{FW}, \ac{HW}, \ac{QW}) \\
         Uni &  Unipolarity (if \ac{HW} or \ac{QW}) \\
    \end{tabular}    
    \label{tab:design}
\end{table}

\begin{table}[t!]
    \centering
    \caption{Fixed Parameters}
    \begin{tabular}{c l}
    $f_1$ & Fundamental frequency \\
    $T_s$ & Interlocking time \\
    $L$ & Levels of converter \\
    $q, h$ & Harmonics specifications
    \end{tabular}    
    \label{tab:device}
\end{table}

The OPP problem considered in this work seeks to minimize the load current TDD $(\text{TDD}[I_{\text{load}}])$. Instead of optimizing the TDD directly, we use the normalized load-current energy $\norm{I}_2^2$ as the objective. This is justified because $\norm{I}_2^2$ is proportional to $\text{TDD}[I_{\text{load}}]$ through Parseval's relation, so minimizing one is equivalent to minimizing the other.

Table~\ref{tab:constraints} summarizes typical constraints in the OPP problem. The ``Design'' column is checked if the corresponding constraint can be set by the user, and left blank if the constraint is imposed by external specifications, such as device characteristics, grid codes, or structural properties of the problem.

Table~\ref{tab:design} lists the user-settable attributes from Table~\ref{tab:constraints}. The number of switching angles can be any even integer between $k_{\min}$ and $k_{\max}$ (inclusive). \ac{FW}, \ac{HW}, or \ac{QW} symmetry can be imposed on the voltage signal $u(\theta)$. For \ac{HW} or \ac{QW}-symmetric signals, the waveform is considered \textit{unipolar} if $\forall \theta \in [0, \pi]: \ u(\theta) \geq 0$. Unipolarity can be enforced by restricting the set of allowable switching transitions.

Finally, Table~\ref{tab:device} summarizes properties of the power electronic system. Considering the harmonics constraints and the interlocking angle constraint mentioned in Table~\ref{tab:constraints}, the former can be posed on the Fourier coefficients from~\eqref{eq:fourier_coeff} as $a_0(u, \theta) = 0, \ b_3(u, \theta) \in [-0.01, 0.01]$. The vector $q$ and box-regions $h$ represent the harmonics constraints \eqref{eq:fourier_coeff} for $a$ and $b$ respectively. As an example when $q^a \geq 1$, a sine-harmonic constraint has a representation of via $\textstyle \frac{-1}{\pi} \int_{0}^{2\pi} u(\theta) \cos(q^a \theta) d\ \theta \in h^a$. In this manner, equality and inequality constraints can be imposed on specific harmonics $(a, b)$ of a pulse pattern $u(\theta)$. The interlocking angle $\Theta$ is based on an interlocking time $T_s$ provided by the device manufacturer. This time (e.g., $50\,\mu$s) is the minimum duration required between any two switching transitions to prevent short circuits in the dc-link capacitors of the converter, and is related to the fundamental frequency as $\Theta = T_s f_1$.

\subsection{Optimal Pulse Pattern Problem}

For a fixed number of switching angles $k$, the $k$-switch-restricted OPP synthesis problem involves the optimization variables $\{\alpha_{i}\}_{i=1}^k$ (switching angles) and $\{n_{i}\}_{i=0}^k$ (levels of the switching signal). The variables $n_i$ are indices corresponding to the levels $u_i = u_{n_i}$. Parameterizing in terms of $n_i$ rather than directly using $u_i$ allows searches over inhomogeneous levels \cite{dahidah2014review}, such as $L = \{-1, -2/3, 0, 2/3, 1\}$. The minimal-energy OPP synthesis problem with $k$ switching angles is as follows:
\begin{prob}
    Given the system properties in Table \ref{tab:device}, $k$ switching angles, and requirements (Sym, Uni),  determine a switching sequence ($\{\alpha^i\}, \{n^i\}$) to minimize:
    \label{prob:tdd_orig}
\begin{subequations}
    \label{eq:tdd_orig_alpha}
\begin{align}
    J^*(k) = &\min_{\alpha^i, n^i} \norm{I}_2^2 \label{eq:tdd_orig_obj}\\
   \text{s.t.}\quad & \text{Harmonics} (q, h; \alpha, u_n)
    \label{eq:tdd_orig_alpha_she} \\
    & n^{i+1} - n^i \in \{-1, 1\} & & \forall i \in 0..k-1 \label{eq:tdd_orig_step}\\
    & \alpha^{i+1} \geq \alpha^i + \Theta &  &\forall i \in 1..k-1\label{eq:tdd_orig_alpha_lock}\\ 
    & \alpha_1 - \alpha_{k} - 2\pi \geq \Theta \label{eq:tdd_orig_alpha_lock_end} \\
    & n^i \in 1..N & &  \forall i \in 0..(k-1) \label{eq:tdd_orig_alpha_L}\\
    & n^0 = n^k \label{eq:tdd_orig_alpha_n_end} \\    
    & \text{Symmetry and Unipolarity}.
\end{align}
\end{subequations}
\end{prob}
The main sources of nonconvexity in Problem~\ref{prob:tdd_orig} are the integer-valued variables~\eqref{eq:tdd_orig_alpha_L}, the transcendental harmonics constraints~\eqref{eq:tdd_orig_alpha_she}, and the nonlinearities and multiplications in the objective fucntion~\eqref{eq:tdd_orig_obj}. When the harmonics constraints fix the fundamental coefficients $a_1, b_1$ via equality constraints, an optimal pattern $(\alpha, n)$ for $u_{\text{ext}}(\theta)=0$ is also optimal for $\tilde{u}_{\text{load}}(\theta)=A \cos(\theta + \phi)$, since the TDD objective ignores the fundamental component.



%% file: sections/opp_as_ocp.tex
\section{OPP as Mode-Selecting Optimal Control}
\label{sec:opp_as_ocp}
The OPP Problem \ref{prob:tdd_orig} is a static and finite-dimensional optimization problem over the level indices $n$ and the switching angles $\alpha$. This section explores how Problem \ref{prob:tdd_orig} can be mapped into an equivalent and nonconservative dynamical optimization problem, involving mode-selecting optimal control of a hybrid dynamical system. Discussion will focus on the case of \ac{FW} symmetry. Modifications required to apply \ac{HW} or \ac{QW} with possible unipolarity are detailed in Appendix \ref{app:sym_unipolar}.

\subsection{Assumptions}

The following assumptions are imposed for the hybrid-system construction:
\begin{assum}
    The maximal number of switching angles $k_{\max}$ is an even integer.
    \label{assum:switch}
\end{assum}
\begin{assum}
    The interlocking angle $\Theta$ is positive.
    \label{assum:interlock}
\end{assum}
\begin{assum}
There are a finite number of levels in the converter, and  $L$ is sorted in increasing order.
\label{assum:level}
\end{assum}
\begin{assum}
The vectors $q^a$ and $q^b$ are finite-dimensional and have bounded entries (if not empty).
 \label{assum:harmonic}
\end{assum}

These assumptions contribute to posing a well-defined \ac{OPP} problem instance. Specifically, the combination of 
Assumptions \ref{assum:switch} and \ref{assum:interlock} together ensure that there is no   Zeno execution (no infinite number of switches in a finite time) \cite{zhang2001zeno}. Assumption \ref{assum:level} is a finite-set constraint on the inverter topology.  Assumption \ref{assum:harmonic} implies that only a finite number of finite-frequency harmonics are constrained. 

\subsection{Hybrid System Description}

The hybrid system will be defined with respect to a sequence of modes indexed by $(n, i)$, where $n \in 1..N$ is the index for the level and $i \in 0..k$ is the number of previously elapsed switching transitions. Each mode of the hybrid system will follow linear dynamics based on \eqref{eq:dynamics_per_unit}, differing only in the applied voltage $u_n$. The key insight used developing  the hybrid system formulation is the equivalence between the following two concepts:
\begin{enumerate}
    \item  A switching angle-level pair $(\alpha^i, u_{n^i})$.
    \item A dwell time of $\alpha^{i+1} - \alpha^i$ spent in mode $(n, i)$.
\end{enumerate}


The dynamics will be expressed in a trigonometric framework with respect to variables $(c, s)$ modeling $(\cos(\theta), \sin(\theta))$ (as performed in SHE), in which the variables $(c, s)$ traverse counterclockwise around the unit circle at 1 radian/time unit. Such a polynomial reformulation of the angle $\theta$ into $(c, s)$ will ensure that the resultant hybrid system OCP may be approximated through polynomial-optimization-based convex relaxation techniques. 

\subsubsection{Transition Graph }

We begin by describing a transition graph $\gs$ that represents the set of possible switching transitions. The vertices $\vs$ of the transition graph are 
\begin{align}
\label{eq:vertices}
    \vs = \left\{(n, i) \mid \begin{array}{ll}
         n \in 1..N, \ \  i \in 0..k \\        
    \end{array} \right\}.
\end{align}
The vertices are doubly indexed by the current level $n$, and by the number of elapsed switching transitions since the beginning of the fundamental period $i$.

The edges $\es$ of the transition graph are formed by step-up and step-down edge sets $\es^\pm$:
\begin{align}
    \es^+ &= \{(n, i) \rightarrow (n\pm 1, i+1), n \in 1..N-1,\   i \in 0..k-1\} \nonumber\\
    \es^- &= \{(n, i) \rightarrow (n\pm 1, i-1), n \in 2..N  i \in 0..k-1\} \nonumber\\ 
    \es &= \es^+ \cup \es^-. \label{eq:edges}
\end{align}
The graph $\gs$ as described by \eqref{eq:vertices} and \eqref{eq:edges} has a total of $N (k+1)$ vertices and $2(N-1)k$ edges.

\begin{figure}
    \centering
    \includegraphics[width=0.7\linewidth]{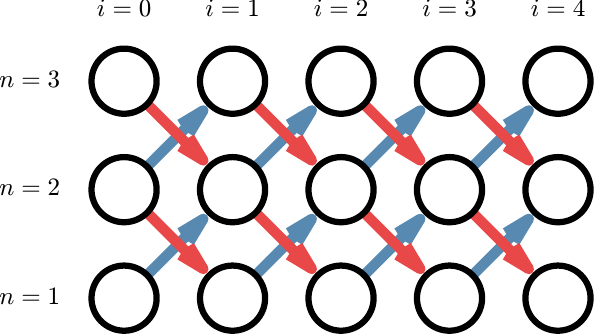}
    \caption{Transition graph for $N=3, k=4$}
    \label{fig:transition_graph}
\end{figure}

Fig.~\ref{fig:transition_graph} displays a transition graph for an $N=3$-level inverter with $k=4$ pulses. The black-bordered circles represent the vertices $(n, i) \in \vs$, which are arranged vertically by $n$ and horizontally by $i$. The red arrows are the step-down transitions in $\es^-$, and the blue arrows are the step-up transitions in $\es^+$.

We denote $\text{Path}^k(\gs)$ as the set of paths in the graph starting at a vertex $(n^0, 0)$ and ending at a vertex $(n^k, k)$. The set of periodic paths $\text{Path}^k_0(\gs) \subset \text{Path}^k(\gs)$ are the subset satisfying the restriction $n^0 = n^k$. 
The chosen sequence of levels $\{n^i\}$ from Problem \ref{prob:tdd_orig} will therefore be represented as a periodic path $P \in \text{Path}_0^k(\gs)$. The switching angles $\alpha^i$ will represent the amount of angle spent in the vertex $(n, i)$.

\begin{figure}[t!]
    \centering
     \begin{subfigure}[b]{0.7\linewidth}
         \centering
         \includegraphics[width=0.8\textwidth]{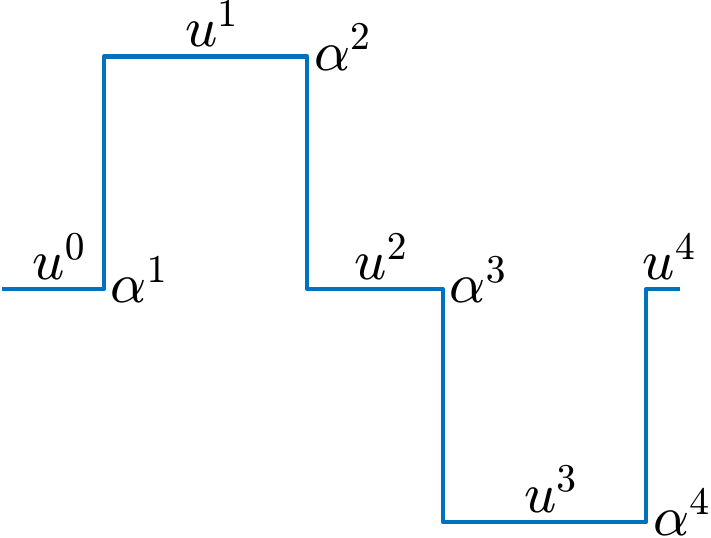}
         \caption{\textrm{Original pulse pattern}}
         \label{fig:association_pulse}
     \end{subfigure}
     \vspace{0.5cm}
     
     \begin{subfigure}[b]{0.7\linewidth}
         \centering
         \includegraphics[width=\textwidth]{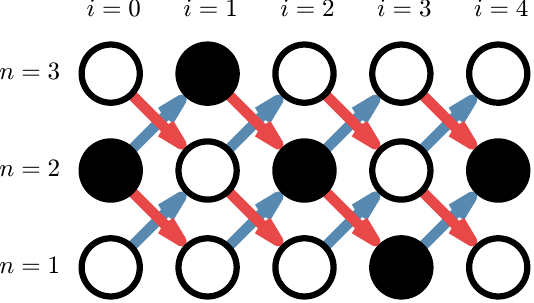}
         \caption{\textrm{ Path representation}}
         \label{fig:association_path}
     \end{subfigure}
          \vspace{0.5cm}

          \begin{subfigure}[b]{0.7\linewidth}         \centering\includegraphics[width=\textwidth]{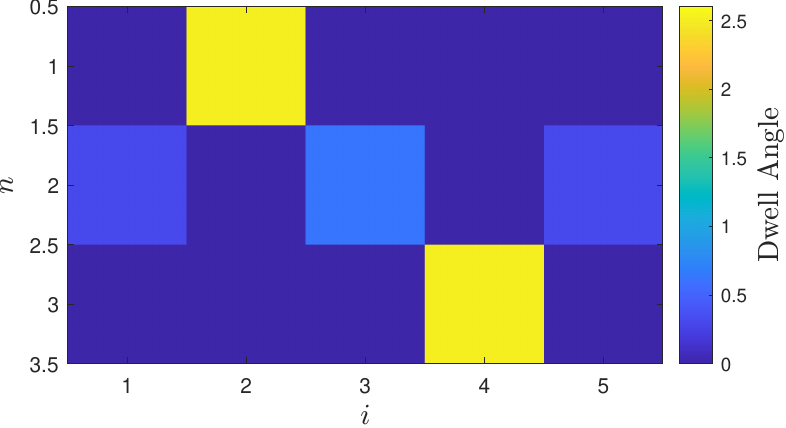}
         \caption{\textrm{ Dwell/Occupancy  table }}
         \label{fig:association_dwell}
\end{subfigure}
              
         \vspace{0.01cm}
    \caption{Representations of a pulse pattern for $N=3, k=4$.}
    \label{fig:transition_association}
\end{figure}

Fig.~\ref{fig:transition_association} visualizes the association between a pulse pattern, a path in a transition graph, and a dwell table for a system with $N=3, k=4$. The pulse pattern in the top-most Fig.~\ref{fig:association_pulse} has a level sequence of $n = [2, 3, 2, 1, 2]$. The middle Fig.~\ref{fig:association_pulse} highlights the vertices involved in this particular path with filled black circles. However, the path representation in Fig.~\ref{fig:association_pulse} only parameterizes the level sequence $\{n^i\}$, and does not incorporate angle information from $\{\alpha^i\}$. The bottom Fig.~\ref{fig:association_dwell} visualizes the combination of $\{n^i\}$ and $\{\alpha^i\}$ as a dwell table, in which each $(n, i)$ mode  is enriched with the angle $\alpha^i$ spent inside the mode (according to the color bar).

\subsubsection{Mode and Switch Dynamics}
\label{sec:support_set}
A linear system will evolve in each mode $(n, i)$ of the transition graph $\gs$. The states at each mode are cosine-angle $c$, sine-angle $s$, clock angle $\phi$, and load current $I$. The states may be accumulated into $x = [c, s, \phi, I]$.
The per-mode dynamics in mode $(n, i) \in \gs$ are  
\begin{align}
    \dot{c} &= -s &  \dot{s} &= c &  \dot{\phi} &= 1 & \dot{I} &= u_i.\label{eq:mode_dynamics}
\end{align}

The dynamics in \eqref{eq:mode_dynamics} vary only in the level $u_i$. Support sets for dynamics will be defined with respect to fixed $k$. If $I_S \subseteq \R$ is an invariant set for the load current $I(\theta)$, then the
the initial condition for dynamics in \eqref{eq:mode_dynamics} are
\begin{align}
    X^0_{n, 0} = \{1\} \times \{0\} \times  [0, 2\pi -\Theta k] \times I_S. \label{eq:set_initial}
\end{align}
Due to \ac{FW} symmetry (2$\pi$-periodicity), the terminal set is the same as the initial set.

Dynamics in \eqref{eq:mode_dynamics} can evolve within the state
\begin{align}
 X_{n, i} &= B([\Theta i, 2\pi - \Theta(k-i)]) \times [0, 2 \pi - \Theta k] \times I_S. \label{eq:set_flow}
\end{align}


The jump dynamics after performing a switch are
\begin{align}
    c_+ &= c &  s_+ &= s &  \phi_+ &= 0 & I_+ &= I.\label{eq:jump_dynamics}
\end{align}
The dynamics in \eqref{eq:jump_dynamics} are the same between for the traversal of any edge in $\es$. The effect of these dynamics are to zero out the clock variable $\phi$. The arc  $e : (n \mp 1) \rightarrow (n, i+1)$ can only be taken if $x(\theta)$ lies within the following guard set:
\begin{align}
    G_{n, i}^\pm = B([\Theta i, 2\pi - \Theta(k-i)]) \times [\Theta, 2 \pi - \Theta k] \times I_s. \label{eq:support_guard}
\end{align}

The interlocking time constraint is enforced due to the presence of the $\phi \geq \Theta$ constraint in \eqref{eq:support_guard}.


\subsubsection{Integration}

The demand distortion objective, harmonics constraint, and power dissipation constraints can be evaluated as accumulations (integrals or sums) over the controlled trajectory of the hybrid system. As an example, the demand distortion penalty $\norm{I}_2^2 = \int_{0}^{2\pi} I(\theta)^2 d \theta$ must be evaluated over all modes traversed by the path.

Let $P \in \text{Path}^k_0(\gs)$ be a periodic path, and let $\{\alpha^i\}$ be a set of switching angles in accordance with the interlocking time constraint. These two variables may be lumped together into an object $\mathcal{T} = (\{\alpha^i\}, \{n^i\})$.
We can define the $\mathcal{T}$-dependent functions $\text{Loc}(\cdot; \mathcal{T}): [0, 2\pi] \rightarrow \mathcal{V}$ and  $\text{Arc}(\cdot; \mathcal{T}): [0, 2\pi] \rightarrow \mathcal{\es}$ using the switching transition index map $i^*(\cdot; \mathcal{T}): [0, 2\pi] \rightarrow 0..k$ as
\begin{align}
i^*(\theta; \mathcal{T}) &= \max_{i \in 0..k} i  \quad \text{s.t.} \ \   \theta \geq \alpha^i\\
    \text{Loc}(\theta; \mathcal{T})&= (n^{i^*(\theta, \mathcal{T})}, i^*(\theta, \mathcal{T})). \\
    \text{Arc}(\theta; \mathcal{T}) &= \begin{cases}
        \varnothing & \theta \not\in \{\alpha^i\} \\
        (n^i, i) \rightarrow (n^{i \pm 1}, i+1) & \theta  = \alpha^i
    \end{cases}
\end{align}
The function $\text{Loc}$ identifies the residing mode at angle $\theta$, and $\text{Arc}$ returns the traversed edge at $\theta$. Due to the interlocking time requirement (Assumption \ref{assum:interlock}), the function $\text{Arc}$ is well-defined given $\mathcal{T}$.
We then define an indicator functions on the modes and the arcs:
\begin{align}
    \chi_{n, i}^\mathcal{T}(\theta) &= \begin{cases}
        1 & \text{Loc}(\theta; \mathcal{T}) = (n, i) \\
        0 & \text{else}
    \end{cases} \\
    \chi_{e}^\mathcal{T}(\theta) &= \begin{cases}
        1 & \text{Arc}(\theta; \mathcal{T}) = e \\
        0 & \text{else}.\end{cases}
\end{align}

We proceed to define integration operators $\Lambda^\vs_\mathcal{T}$, $\Lambda^\es_\mathcal{T}$ over the modes and arcs respectively. Given \ac{FW}-symmetric function $z(x, u)$ (right-continuous with left limits), we define the integration maps $\Lambda^\vs_\mathcal{T}$, $\Lambda^\es_\mathcal{T}$ as 
\begin{subequations}
    \label{eq:integral_operator}
\begin{align}
    \Lambda_\vs^\mathcal{T}: z(x, u) &\mapsto  \sum_{(n, i) \in \vs} \int_{\theta=0}^{2 \pi} \chi_{(n, i)}^\mathcal{T}(\theta)z(x(\theta), u_n) d \theta \\ 
        \Lambda_\es^\mathcal{T}: z(x, u) &\mapsto  \sum_{e \in \es}  \chi_{e}^\mathcal{T}(\theta)z(x(\theta), u_n)
\end{align}
\end{subequations}

The demand distortion objective in \eqref{eq:tdd_orig_obj} can be written as $\norm{I}_2^2 \rightarrow \Lambda_{\vs}^\mathcal{T}[I^2]$. The harmonics constraints can be expressed as integral maps over polynomials in $(c, s)$ using Chebyshev polynomials, as performed in SHE \cite{chiasson2004unified}.
The cosine terms are evaluated using Chebyshev polynomials of the first kind with
\begin{align}
    T_0(c) &= 1, \  T_1(c) = c, \ T_{\ell+1}(c) = 2c T_{\ell}(c) - T_{\ell-1}(c) \nonumber  \\
    \cos(\ell \theta) &= T_{\ell}(\cos(\theta)) \rightarrow T_{\ell}(c),
    \intertext{and the sine terms can be treated using Chebyshev polynomials of the second kind under}
    U_0(c) &= 1, \  U_1(c) = 2c, \ U_{\ell+1}(c) = 2c U_{\ell}(c) - U_{\ell-1}(c) \nonumber  \\
    \sin(\ell \theta) &= \sin(\theta) U_{\ell-1}(\cos(\theta)) \rightarrow s U_{\ell-1}(c).
\end{align}

As an example, a harmonics constraint conversion could be
\begin{align}
    \int_{0}^{2\pi} \sin(5 \theta) u(\theta) d \theta  &\leq 0.01 & \rightarrow & &  \Lambda_\vs^{\mathcal{T}}[s U_{4}(c)] 
\leq 0.01.
\end{align}


\subsection{Hybrid Optimal Control Problem}

Our proposed hybrid system interpretation of Problem \ref{prob:tdd_orig} is the following optimal control task:

\begin{prob}
\label{prob:tdd_hy}
For an inverter control task with device and design requirements from Tables \ref{tab:device} and \ref{tab:design}, choose an initial condition $(\phi_0, I_0)$ and a sequence $\mathcal{T}$ to solve the following problem
\begin{subequations}
    \label{eq:tdd_hy}
\begin{align}
    J^*_{hy}(k) = &\inf_{\phi_0, I_0, \mathcal{T}} \quad \Lambda_\vs^\mathcal{T}[I^2]  \label{eq:tdd_hy_obj}\\
 \text{s.t. } 
 & \text{Harmonics via} \   \Lambda_\vs^\mathcal{T}, \quad\label{eq:tdd_hy_she} \\
    & x(\theta) \text{ follows  \eqref{eq:mode_dynamics} when Loc}(\theta;\mathcal{T}) =  (n, i) \label{eq:tdd_hy_follow} \\
    & x(\theta) \text{ resets as \eqref{eq:jump_dynamics} when } \theta \in \{\alpha^i\} \\    
    & \phi(\theta) \geq \Theta \text{ when jumping (constraint \eqref{eq:support_guard})} \label{eq:tdd_hy_guard}\\
    & \phi(0) =  \phi_0, \  I(0) =  I_0,  \label{eq:tdd_hy_clock}\\    
    & \phi_0 \in [0, 2 \pi - \Theta k],  \ I_0 \in I_S \\
    & \phi(2\pi) = \phi(0), \ I(2\pi) = I(0) \label{eq:tdd_hy_endlock}\\
    & P \in \text{Path}_0^k(\gs) \\
    & \mathcal{T} = (\{\alpha^i\}_{i=}^{k}, P) \label{eq:T_lump}\\
    &\forall i \in 1..k-1: \  \alpha^{i+1} \geq \alpha^{i} \label{eq:tdd_hy_lock}\\    
    & \alpha_1 \geq 0, \ \alpha^k \leq 2\pi \\
    & \text{Symmetry/Unipolarity restrictions on $P$.}
\end{align}
\end{subequations}
\end{prob}

Problems \ref{prob:tdd_orig} and \ref{prob:tdd_hy} have the same objective for fixed $k$. Specifically, a pulse pattern that is feasible for Problem \ref{prob:tdd_orig} admits a representation in $\mathcal{T}$ through a $u \rightarrow P$. While a generic $\mathcal{T}$ may violate the interlocking time requirement, obedience of the interlocking time is assured by constraint \eqref{eq:tdd_hy_guard}. 


\subsection{Convex Relaxation}

We apply moment-based convex relaxation techniques to acquire lower-bounds on the TDD objective of Problem \ref{prob:tdd_hy}.  Refer to Appendix \ref{app:measure} for technical preliminaries about Borel measures and moment-SOS methods, and to Appendix  \ref{app:measure_program} for the specific formulation of the OPP convex relaxation.

The convex relaxation may be interpreted through the lens of dwell/occupancy tables as used in Markov Decision Processes \cite{puterman1990markov}. Given a transition graph $\gs$ with vertices $(n, i) \in \vs$, we define a \textit{dwell table} as a set of $\vs$-indexed dwell angles $\xi_{v} \in \R^{\abs{\vs}}$ such that 
\begin{align}
    \xi_{v} &\geq 0, &  \textstyle  \sum_{v \in \vs} \xi_{v} &= 2\pi.    \label{eq:dwell_constraint_base}
\end{align}

The dwell angles $\xi_v$ are interpreted both as the angular difference spent in a mode, and as the mass of a Borel measure that describes the pulse pattern trajectory. The set of all dwell tables $\{\xi_v\}_{v \in \vs}$ is convex. 

A pulse pattern $\mathcal{T} = (\alpha, n)$ can be mapped into a \textit{pure dwell table} in a one-to-one fashion:
\begin{subequations}
\begin{align}
    \xi_{(n, 0)} &=  \begin{cases}
    \alpha^1 & n = n^0 \\
        0 & \text{else}
    \end{cases} \\
    \forall i \in 1..k-1: \ \xi_{(n, i)} &=  \begin{cases}
    \alpha^{i+1} - \alpha^i & n = n^i \\
        0 & \text{else}
    \end{cases} \\
    \xi_{(n, k)} &=  \begin{cases}
    2\pi - \alpha^k & n = n^k \\
        0 & \text{else}.
    \end{cases}
\end{align}
\end{subequations}

The interlocking time $\Theta$ constraint induces a further restriction on admissible dwell tables:
\begin{subequations}
\label{eq:dwell_constraint_interlock}
\begin{align}
    \forall i \in 1..k-1: \qquad \sum_{n \mid (n, i) \in \vs} \xi_{(n, i)} &\geq \Theta \\
     \sum_{n \mid (n, 0) \in \vs} \xi_{(n, i)} + \sum_{n \mid (n, k) \in \vs} \xi_{(n, i)} &\geq \Theta.
\end{align}
\end{subequations}
Fig.~\ref{fig:association_dwell} plots a pure dwell table arising from the pulse pattern in Fig.~\ref{fig:association_pulse}. Properties of pure dwell tables include
%
\begin{subequations}
\begin{align}
\xi_{(n, i)} > 0 &\implies \xi_{(n', i)} = 0 \qquad \forall n'\neq n \\
\xi_{(n, i)} > 0 &\implies \xi_{(n, i+1)} = 0   \\
\xi_{(n, i)} > 0 &\implies \xi_{(n+1, i+1)} > 0 \ \text{XOR} \ \xi_{(n-1, i+1)} > 0.
\end{align}
\end{subequations}

\begin{figure}
    \centering
    \includegraphics[width=\linewidth]{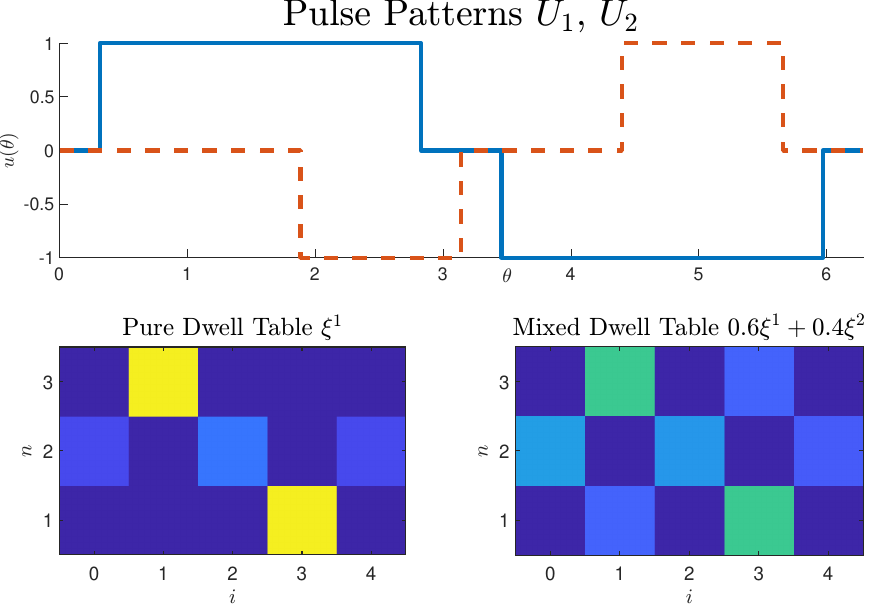}
    \caption{A pure and a mixed dwell table for $k=4$}
    \label{fig:dwell_pure_mixed}
\end{figure}

A \textit{mixed dwell table} is a dwell table that is constructed from a convex combination of dwell tables. The top pane of Fig.~\ref{fig:dwell_pure_mixed} plots pulse patterns $U_1$ (blue, solid) and $U_2$ (orange, dashed). The two pulse patterns have respective dwell tables $\xi^1$ and $\xi^2$. The bottom-left pane of Fig.~\ref{fig:dwell_pure_mixed} plots the dwell table $\xi^1$ of  $U_1$, while the bottom-right pane plots the mixed dwell table from the convex combination  $\xi^{\text{mix}} = 0.6 \xi^1 + 0.4 \xi^2$.

\begin{algorithm}[h]
    \caption{Approximate extraction of pulse patterns from dwell tables}\label{alg:algorithm1}
      \KwData{Transition graph $\gs$, Dwell table $\xi$.} 
  \KwResult{Recovered pattern $\mathcal{T} = (n, \alpha)$}
    $n^0 = \argmax_{n \in 1..N} \xi_{(n, 0)} $ \;
    \For{i $\in 1..k$ }{
    $n^i = \argmax_{n \in \{n^{i-1} - 1, n^{i-1} + 1\}} \xi_{(n, i)}$ \;
    }
    $\Xi = \sum_{i=0}^k \xi_{(n^i, i)}$ \;
    \For{i $\in 1..k$ }{
    $\alpha^i = \frac{2\pi}{\Xi} \sum_{j=0}^i \xi_{(n^{j}, j)}$ \;
    }
\end{algorithm}

Given a dwell table $\xi$, a pulse pattern $\mathcal{T}$ can be approximately extracted by the greedy procedure described in Algorithm \ref{alg:algorithm1}.
If the input dwell table $\xi$ is a pure dwell table arising from a pulse pattern $\mathcal{T}$,  Algorithm \ref{alg:algorithm1} will exactly recover $\mathcal{T}$. In the case where $\xi$ is a mixed dwell table, Algorithm \ref{alg:algorithm1} will recover \textit{a} pulse pattern $\mathcal{T}^\text{rec}$, but this pulse pattern may not be feasible for the problem constraints (e.g. harmonics, interlocking time) of Table \ref{tab:constraints}. The recovered pulse pattern $\mathcal{T}^\text{rec}$ may then be used as the initial seed for a local search (e.g. \texttt{fmincon}); optimizing over $\alpha$ with respect to a fixed level sequence $n$.


The dwell table $\xi$ possesses is 0-th order information of a pattern $\mathcal{T}$, containing only the occupancy times in $\gs$.
The harmonics constraints and TDD cost require higher order information, corresponding to moments of the Borel measures with degree $\geq 1$. The convex approximation of Appendix \ref{app:measure_program} is computationally realized by a sequence of semidefinite programs in increasing size, parameterized by the polynomial degree $\beta$. The variables of these semidefinite programs \cite{lasserre2009moments} are `pseudo-moments,' or real numbers that satisfy necessary conditions to be the moments of measures arising from an admissible pulse pattern. The dwell table $\xi$ are a subset of the variables in this semidefinite program, corresponding to 0-th order psuedo-moments of occupancy measures. 
These necessary conditions are tightened as the polynomial degree $\beta$ increases, thus yielding a rising sequence of lower bounds to the true minimal TDD.

The time complexity of solving the semidefinite programs scales in the worst case as $O((\abs{\vs} + \abs{\es})^{3/2} \beta^{16})$ \cite{claeys2016modal}. In the case of \ac{FW} symmetry with $\abs{\vs} = N (k+1)$ and $\abs{\es} = 2 N k$, the computational impact of $O((Nk)^{3/2} \beta^{16})$ is jointly sub-quadratic scaling in $N$ and $k$.

Given a dwell table $\xi$ derived from the solution of this SDP, application of Algorithm \ref{alg:algorithm1} yields a pulse pattern $\mathcal{T}$. Subsequently performing a local search starting from this recovered $\mathcal{T}$ may yield either a feasible pattern or return infeasibility. If the local search returns a feasible pattern $\mathcal{T}_\text{feas}$, there is no guarantee that $\mathcal{T}_\text{feas}$ minimizes the TDD. Instead, the TDD generated by $\mathcal{T}_\text{feas}$  can be compared to the TDD lower bound computed by the SDP in order judge suboptimality of $\mathcal{T}_\text{feas}$ or other candidate patterns.




%% file: sections/examples.tex
\section{Numerical Examples}

\label{sec:examples}


Numerical experiments were conducted in MATLAB (R2024a). The  required include GloptiPoly 3 \cite{henrion2009gloptipoly} (generating the moment relaxations), YALMIP \cite{lofberg2004yalmip} (parsing the programs), and Mosek \cite{mosek110} (solving the convex optimization problems via primal-dual interior point methods). Code to replicate these examples is publicly available \footnote{\url{https://github.com/jarmill/opp_pop}}.

All numerical examples in this section will involve control of a five-level converter with levels $L = \{
        -1, \ -0.5, \ 0, \ 0.5,  \ 1\}$.
The system has a fundamental frequency of $f_1=50$ Hz and an interlocking time of $T_s = 100\,\mu$s, yielding an interlocking angle of $\Theta = \pi/100 = 0.0314$ rad. The applied voltage is $\tilde{u}_{\text{load}}=0$. 
The modulation index $M$ in the design requirements for a pulse pattern is the desired value of the first harmonic $b_1$.

\subsection{Single Pattern}

\label{sec:single_pattern}

This case involves a load with ratio $\tau = R_{\text{load}}/L_{\text{load}} = 0.5$. A \ac{QW}-symmetric pulse pattern with $k=24$ and first harmonic $b_1=0.8$ must be synthesized to minimize the current TDD. No external voltage signal nor power budget is applied to this problem. The reference load current is
\begin{align}
    I^*(\theta) & = (1.6/\sqrt{5}) \cos(\theta + \tan^{-1}(0.5)).
\end{align}

The only applied harmonics constraint outside of QW symmetry is $b_1 = 0.8$. There are 16 unique paths through the transition graph $\gs$ $(i=0 \rightarrow i = 6)$ under the unipolarity and QW impositions.

Solving a convex relaxation of Problem \ref{prob:tdd_hy} with polynomial order $\beta = 3$ under 2 partitions ($[0, \pi/2] \cup [\pi/2, \pi]$) leads to a lower-bound $p^*_3 = 1.6092$ on $\norm{I}_2^2$. Direct extraction of a pulse pattern from the SDP solution by Algorithm \ref{alg:algorithm1} leads to infeasibility of the harmonics constraints ($b_1 = 0.8205$ rather than the desired $b_1 = 0.8$). Warm starting both \texttt{fmincon} and \texttt{IPOPT} at a fixed sequence $u$ leads to a candidate pulse pattern with $\norm{I}_2^2 = 1.6092$ with $\norm{I}_2^2 - p_3^* = 2.151 \times 10^{-5}$ described by parameters
\begin{align}
    u_{0:6} &= (0, 0.5, 1, 0.5, 1, 0.5, 1) \\
    \alpha_{1:6} &= (0.3302, 0.9898, 1.0951,  1.2351, 1.3797, 1.4910). \nonumber
\end{align}
The corresponding lower-bound on the current TDD is 
$\text{TDD} - \text{TDD}_{\text{bound}} =    2.2799\times 10^{-4}$.

\begin{figure}[t!]
    \centering
    \includegraphics[width=\linewidth]{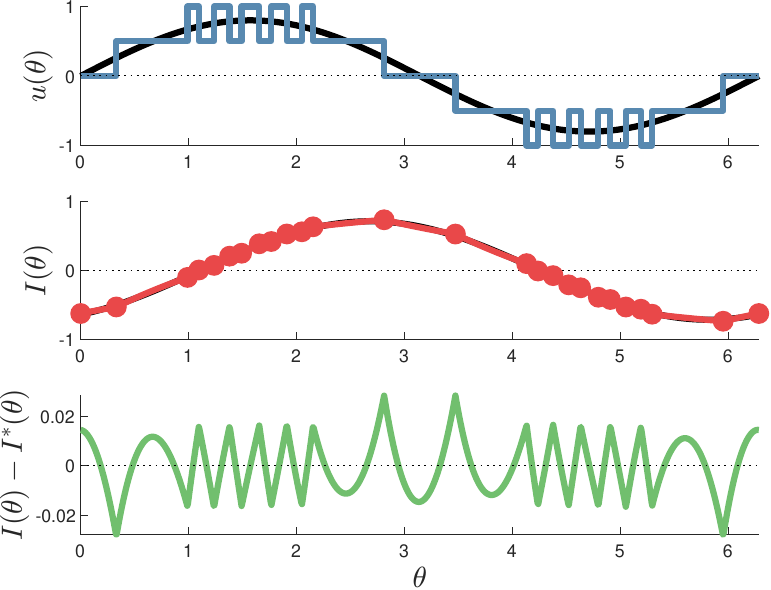}
    \caption{Pattern with 24 switching angles under $\tau = 0.5$}
    
    \label{fig:kappa_5}
\end{figure}

The top pane of Fig.~\ref{fig:kappa_5} visualizes the recovered pulse pattern $u(\theta)$ (blue) as compared to the reference voltage $u^*(\theta) = 0.8 \sin(\theta)$ (black). The middle pane plots the load current $I(\theta)$ (red) and the reference current $I^*(\theta)$ (black). The highlighted red dots in the middle pane are the values of the current $I(\theta)$ evaluated at the switching angles $\alpha$. The bottom pane displays the pointwise difference $I(\theta) - I^*(\theta)$.

We now compare against other approaches.
The polynomial system associated with SHE (QW symmetry with $b_1 = (4/\pi) 0.8, b_{3, 5, \ldots, 11} = 0$) is symmetry-reduced using power sums and elementary symmetric polynomials \cite{yang2016unified, wang2022application} and then solved using a Julia-based Homotopy Continuation method \cite{sommese2005numerical, breiding2018homotopycontinuation}. 
The McCormick-based relaxation scheme EAGO.jl fails to generate feasible solutions for this problem setting with current TDD \cite{eago2022} (but generates candidate solutions with voltage TDD objectives). 

SHE is successfully performed with $k\in\{8, 12, 16, 28\}$ switching angles, and fails at $k\in\{4, 20, 24\}$ angles. At each of the successful $k$ values, only a single path results in a feasible SHE solution.   The $k=28$ case generates a pulse pattern with energy $\norm{I_{\text{SHE}}}_2^2 = 1 + 5.950 \times 10^{-3}$. Solving the order-3 relaxation in \ref{prob:tdd_mom} yields a lower-bound of  $\norm{I}_2^2 \geq 1 + 5.812 \times 10^{-3}$. Recovery of a pulse pattern from the order-3 solution via Algorithm \ref{alg:algorithm1} followed by subsequent \texttt{fmincon} local search
leads to an energy of $\norm{I_{\text{rec}}}_2^2 = 1 + 1.5873 \times 10^{-3}$. Both the SHE and recovered pulse patterns have the same modulation sequence of 
\begin{align}
    u &= (0, 0.5, 0, 0.5, 1, 0.5, 1, 0.5)
    \intertext{with switching angles of}
    \alpha_{\text{rec}} &= (0.2307, 0.3260,    0.4155,
    1.0027,
    1.1413,
    1.3055,
    1.5022) \nonumber\\
    \alpha_{\text{SHE}} &= (    0.1963,
    0.2993,
    0.4225,
    0.9551,
    1.0809,
    1.2748,
    1.4914).\nonumber
\end{align}The difference between the energy values is $\norm{I_{\text{SHE}}}_2^2 - \norm{I_{\text{rec}}}_2^2 = 7.7491\times 10^{-5}$.





\subsection{Parameter Sweep: Length vs. Degree}

\begin{figure}[t!]
    \centering
    \includegraphics[width=0.7\linewidth]{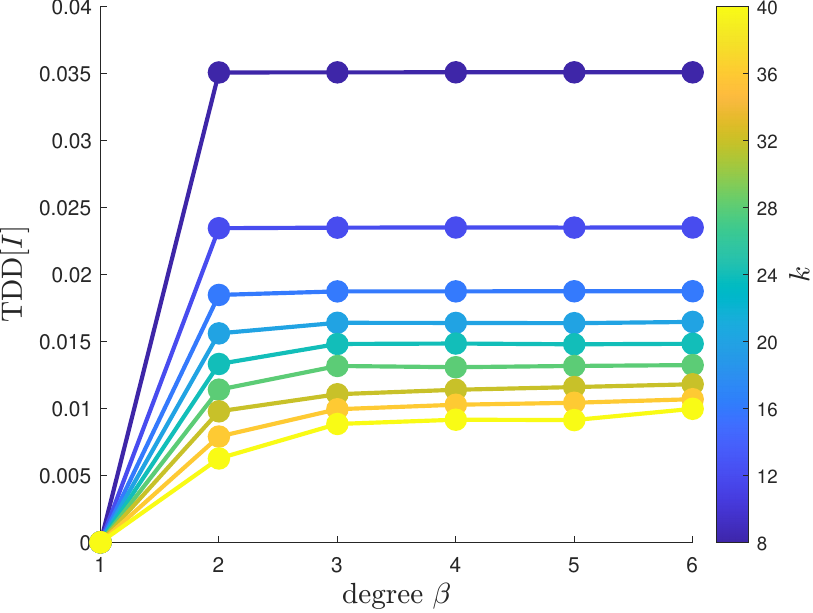}
    \caption{TDD bound vs. degree $\beta$ and length $k$ at $\tau = 1$}
    \label{fig:length_tdd}
\end{figure}

We now change parameters associated with the five-level converter in the previous example, and show how these modifications affect lower bounds on the current TDD.
We first fix a load ratio of $\tau = 1$, with a reference energy of $\norm{I^*}_2^2 = 1.0053$. Fig.~\ref{fig:length_tdd} plots TDD lower-bounds computed by degree-$\beta$ truncations in \eqref{eq:tdd_mom} as $\beta$ ranges from $1..6$ and as $k$ increases from $8..40$. The signal $u(\theta)$ is restricted to be HW-symmetric.

\begin{figure}[t!]
    \centering
    \includegraphics[width=\linewidth]{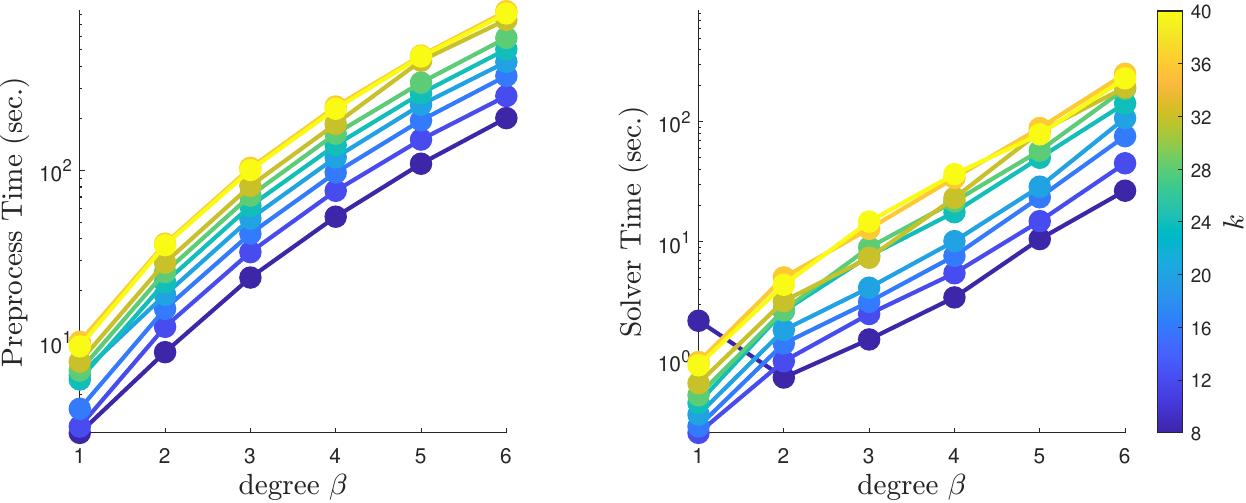}
    \caption{Computation time to produce the result in Fig.~\ref{fig:length_tdd}}
    \label{fig:length_tdd_time}
\end{figure}

The TDD lower bounds in this example obey a monotonicity property: they rise for fixed $k$ with increasing $\beta$, and they fall with fixed $\beta$ and increasing $k$. Fig.~\ref{fig:length_tdd_time} reports the time taken in preprocessing (GloptiPoly and MATLAB overhead) and solving (Mosek execution) the degree-$\beta$ LMI programs.

\subsection{Parameter Sweep: Load Ratio vs. Degree}

\begin{figure}[t!]
    \centering
    \includegraphics[width=\linewidth]{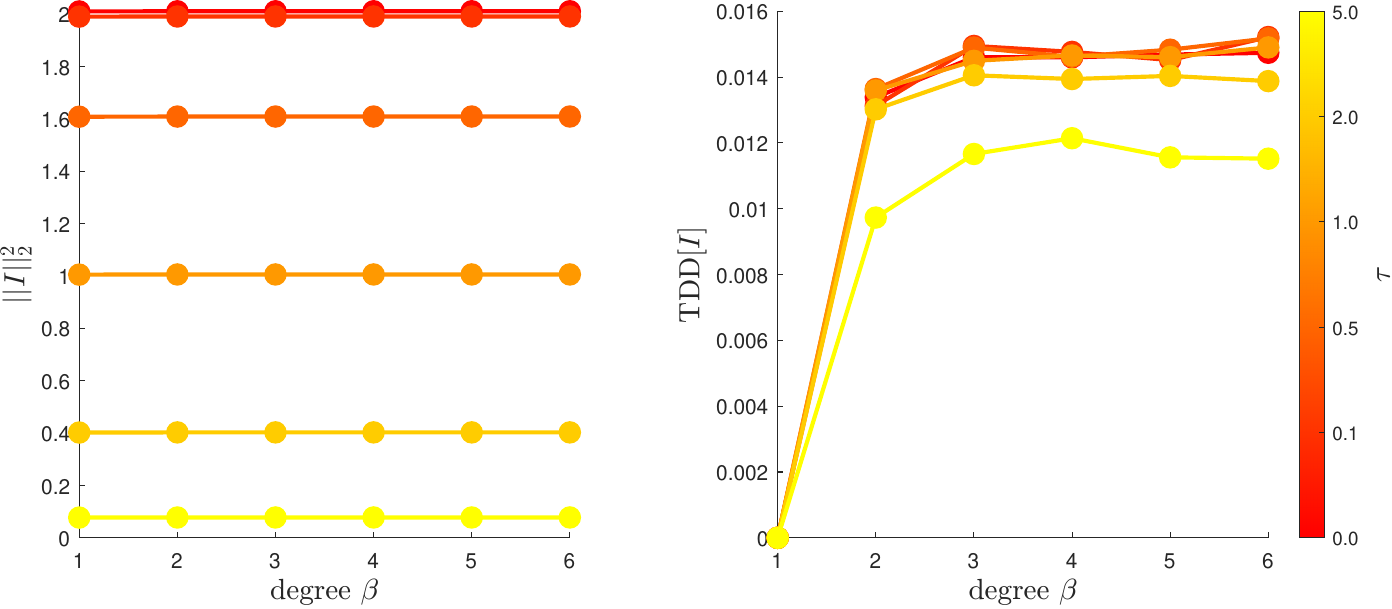}
    \caption{Energy and TDD vs. increasing $\tau$}
    \label{fig:load_kappa}
\end{figure}

\begin{figure}[t!]
    \centering
    \includegraphics[width=\linewidth]{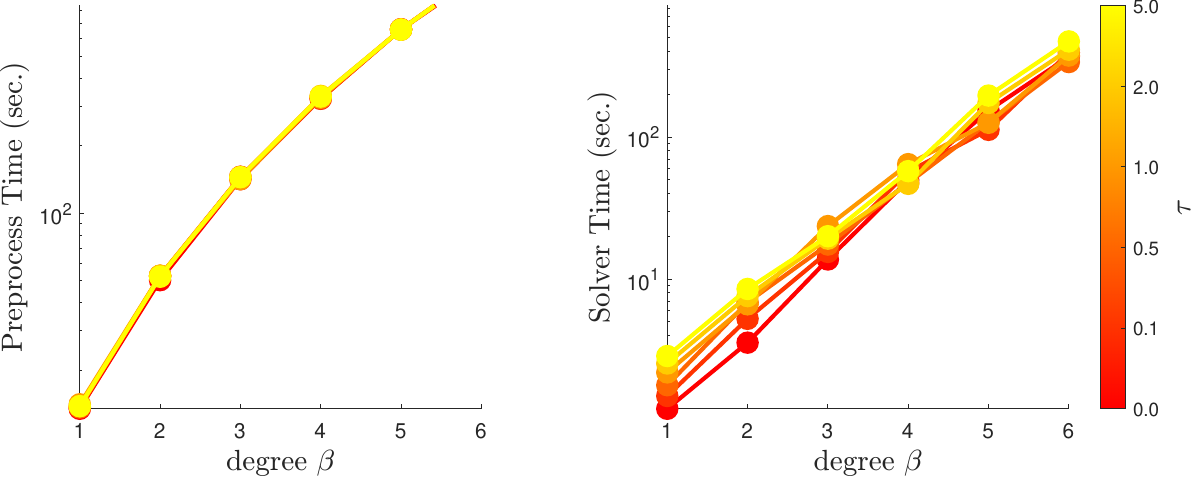}
    \caption{Time to compute Fig.~\ref{fig:load_kappa}}
    \label{fig:load_kappa_time}
\end{figure}

Our second parameter sweep includes increasing the load ratio $\tau$. This scenario involves the same five-level converter with $k=24$ switching transitions at a modulation index $b_1 = M= 0.8$ under the further harmonics constraint that $b_3 \in [-0.1, 0.1]$. SHE is infeasible for this particular scenario, as mentioned in Section \ref{sec:single_pattern}. Fig.~\ref{fig:load_kappa} plots lower-bounds on the signal energy (left) and the TDD (right) as $\tau$ changes from $0$ to $5$. Fig.~\ref{fig:load_kappa_time} displays the preprocessing and computation time required to find these lower-bounds.

\subsection{Parameter Sweep: Modulation Index vs. Length}

\begin{figure}[t!]
    \centering
    \includegraphics[width=\linewidth]{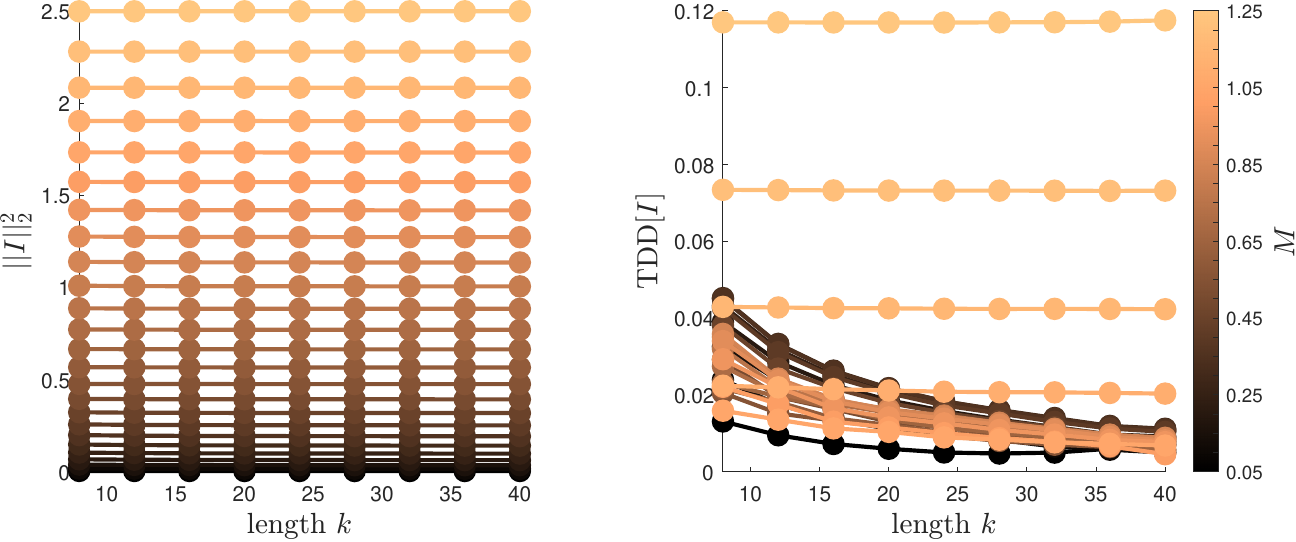}
    \caption{Energy and TDD vs. increasing $k$}
    \label{fig:modlength_tdd}
\end{figure}

\begin{figure}[t!]
    \centering
    \includegraphics[width=\linewidth]{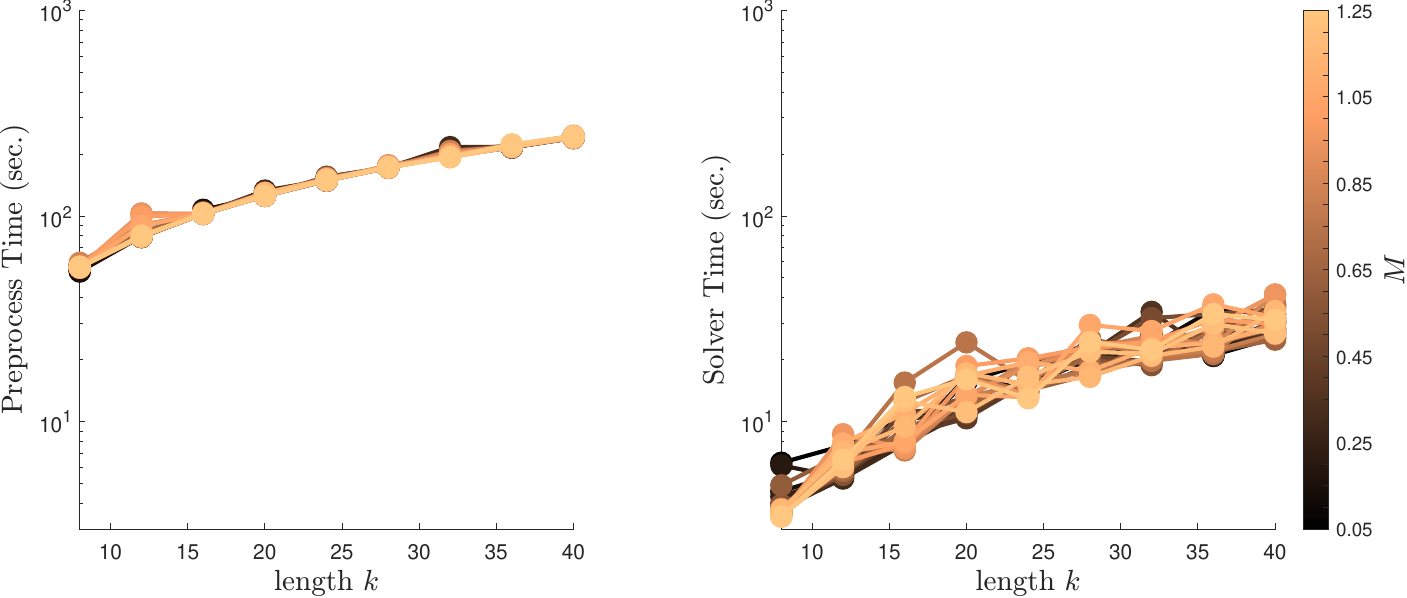}
    \caption{Computation time to produce Fig.~\ref{fig:modlength_tdd}}
    \label{fig:modlength_tdd_time}
\end{figure}

Our final parameter sweep involves fixing the degree and load ratio to $\beta=3$ and $\kappa = 1$. The modulation index is swept from 0.05 to 1.25 with increments of 0.05, and the number of transitions is swept from 8 to 40 with increments of 4. Problem \ref{prob:tdd_mom} is once again solved under HW symmetry with a partition of $[0, \pi/2] \cup [\pi/2, \pi]$. Fig.~\ref{fig:modlength_tdd} plots the energy and TDD lower bounds as $k$ and $M$ increase. The TDD falls as $k$ increases. The TDD is high in the overmodulation regime of $M > 1.155$ \cite{holmes2003pulse}. Fig.~\ref{fig:modlength_tdd_time} reports the preprocessing and solver time to produce the bounds in Fig.~\ref{fig:modlength_tdd}.

\begin{figure}[t!]
    \centering
    \includegraphics[width=\linewidth]{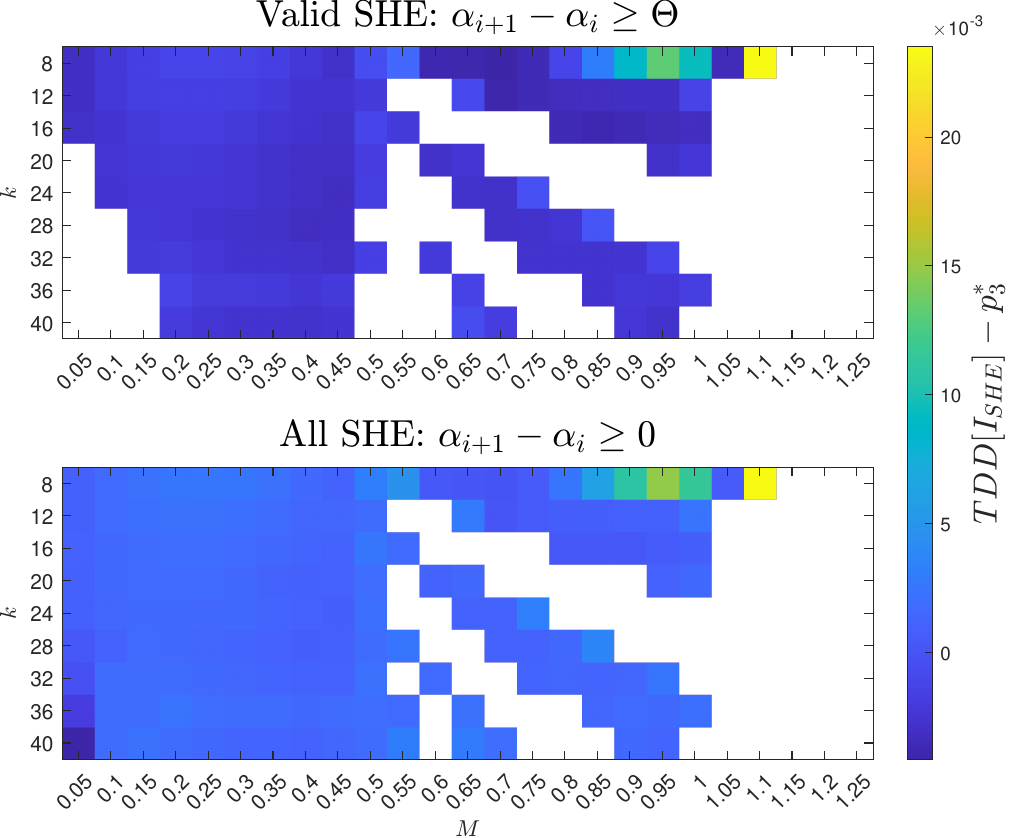}
    \caption{Order $\beta=3$ bounds v.s. SHE}
    \label{fig:she_sweep}
\end{figure}

Fig.~\ref{fig:she_sweep} compares the TDD of SHE-produced pulse patterns to the $\beta=3$ TDD lower bounds from Figure \ref{fig:modlength_tdd}. The top panel from Fig.~\ref{fig:she_sweep} demonstrates that the feasible SHE solutions always have a higher TDD value than the lower bounds: the minimal TDD gap between the SHE and the $\beta=3$ lower bound occurs at $M = 0.9, k=24$ with $\text{TDD}[I_{SHE}] - p^*_3 =  5.7658 \times 10^{-6}$. The bottom panel of Fig.~\ref{fig:she_sweep} displays the TDD gaps associated with all SHE solutions, including the solutions that violate the interlocking angle constraint of $\Theta$. As an example, the $M = 0.05, k=40$ SHE-computed pulse pattern has a minimal angle difference of $\alpha_2 - \alpha_1 = 0.0080$ rad, which is less than the $\Theta = 0.0314$ requirement. The white colored squares in the bottom panel of Fig.~\ref{fig:she_sweep} correspond to parameter settings where SHE fails to find a pulse pattern meeting the imposed harmonics equality constraints. The white colored squares on the top panel either arise from failure of SHE to find a solution, or failure of SHE generate a harmonics-feasible pulse pattern that meet the $\Theta = 0.0314$ constraint.

From each SDP solution associated with a $(k, M)$ pair at $\beta=3$, we use Algorithm \ref{alg:algorithm1} to produce a candidate pulse pattern. We fix the levels $u$ of these patterns, and optimize the angles $\alpha$ using \texttt{fmincon} to minimize the current TDD under constraint feasibility. The local search with initial points of Algorithm \ref{alg:algorithm1} is feasible for all $(k, M)$ except for $(28, 1)$ and $(40, 1.25)$: for these infeasible points we begin the local search at the initial points produced by Algorithm \ref{alg:algorithm1} at $(28, 0.95)$ and $(40, 1.20)$ respectively to obtain a feasible pulse pattern.

\begin{figure}[t!]
    \centering
    \includegraphics[width=\linewidth]{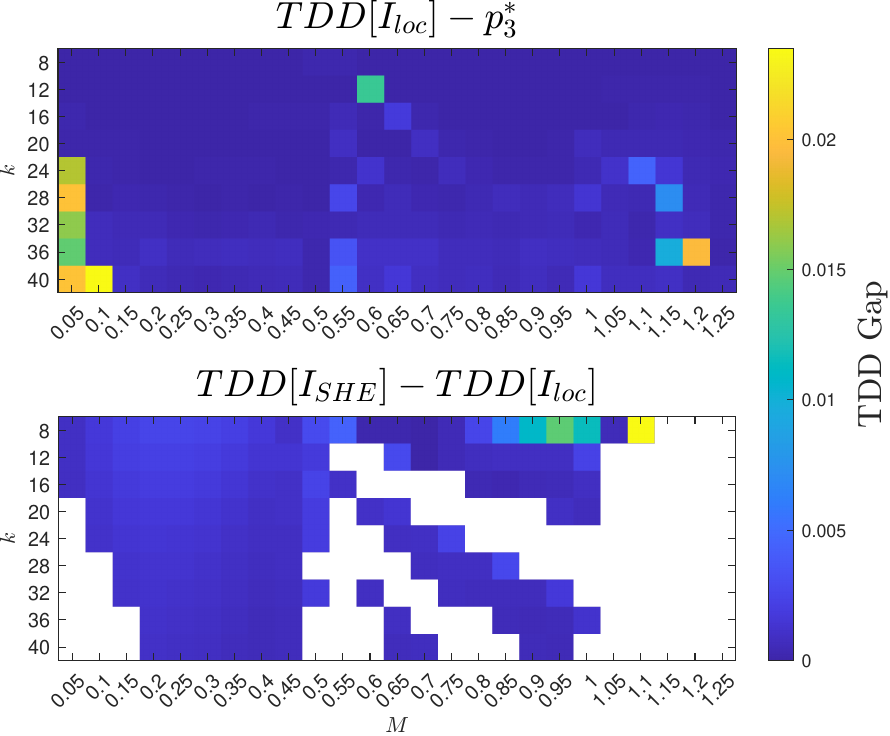}
    \caption{SHE, local search, and $\beta=3$ bounds}
    \label{fig:loc_she}
\end{figure}

The top pane of Fig.~\ref{fig:loc_she} plots the nonnegative gap between the TDD produced by the local search and the $\beta=3$ bound. The bottom pane of Figure \ref{fig:loc_she} plots the difference between the TDD produced by SHE and by the local search. The SHE TDD is less than the local TDD only at $(k, M) = (8, 0.7)$ with a gap of $2.2913\times 10^{-6}$. In all other cases in this experiment, the local search TDD is less than the SHE TDD.





%% file: sections/conclusion.tex
\section{Conclusion}

\label{sec:conclusion}

OPPs are an optimal modulation technique that can achieve favorable steady-state behavior in power electronic systems. While optimizing the pulse patterns reduces harmonic distortions, the optimization problem is highly nonconvex and thus computationally challenging. In this work, we bounded the single-phase OPP problem by first casting it as a periodic mode-selecting OCP of a hybrid system without introducing conservatism. The OCP was then relaxed using established convex methods from optimal control. Applying the moment-SOS hierarchy to this mode-selecting OPP yielded a sequence of semidefinite programs whose optimal values lower-bound the true minimal achievable distortion under device and design constraints. These lower bounds provide a benchmark to assess the optimality of a given pulse pattern.  

Future work will focus on generalizing the single-phase framework to three-phase converters, including bounding differential-mode harmonic distortion under 120$^\circ$ symmetry and imposing common-mode voltage constraints. Additional extensions include incorporating constraints on the power budget arising from the switching and conduction losses of the semiconductor devices~\cite{geyer2023optimized}. On the computational side, efforts will aim to reduce the complexity of the semidefinite programs generated by the moment-SOS hierarchy for OPPs.


%% file: sections/acknowledgements.tex
\section*{Acknowledgements}

The authors would like to thank Didier Henrion, Jie Wang,  Corbinian Schlosser, Irina Suboti\'{c}, Georgios Papaforitou, Mircea Lazar, Cristobal Gonzales, and Carsten Scherer for discussions about \acp{OPP} and convex relaxations.


%% file: appendix/app_energy.tex
\section{Signal Energy Explicit Expression}
\label{app:energy}
This section lists formulas for the signal energy $\norm{I}_2^2$. If $I_p(\theta)$ is the current response of system \eqref{eq:dynamics_per_unit} under the pulse pattern $u(\theta)$ and $I_\text{ext}$ is the current response under the external signal $-u_{\text{ext}}$, then the total signal energy is
\begin{subequations}
\begin{align}
    \norm{I}_2^2 &= \textstyle\int_{0}^{2\pi} (I_\text{ext}(\theta) + I_p(\theta))^2 d \theta \\
    &=\textstyle \norm{I_{\text{ext}}}^2_2 + 2\int_{0}^{2\pi} I_p(\theta)I_\text{ext}(\theta)I_\text{ext}(\theta)  d \theta  + \norm{I_p}^2_2. \label{eq:energy_sum_label}\\
    \intertext{We can label the individual energy terms in \eqref{eq:energy_sum_label} as}
   \norm{I}_2^2 &= E_{\text{ext}} + 2 E_{\text{mix}}  + E_p.\label{eq:energy_total_label}
\end{align}
\end{subequations}

The externally applied signal $u_{\text{ext}}$has a gain of 
\begin{align}
   \gamma_0 &=  A/\sqrt{\tau^2+1} \\
\intertext{and an energy of }
    E_{\text{ext}} &= \pi \gamma_0^2= \pi A^2/(\tau^2 + 1).
\end{align} 

The terms $E_{p}$ and $E_{\text{mix}} $ will be computed in the separate cases of  $\tau=0$ and $\tau \neq 0$. Each case depends on the initial current $I_p(0)$, which we will assume to be given. If the pattern $u(\theta)$ is zero-mean, then $I_p(0)$ can be chosen to generate a zero-mean (minimal energy) current waveform $I_p(\theta)$. Expressions will be written with respect to the switching angle differences $\Delta \alpha^i = \alpha^i - \alpha^{i-1}$ under the extended angle convention $\alpha^0=0$ and $\alpha^{k+1} = 2\pi$. 

\subsection{Pure Reactance}

We first consider the case where $\tau=0$. We define $I_p^i = I_p(\alpha^i)$ as the current produced by the pulse pattern $u(\theta)$ evaluated at the switching angles $\alpha^i$ with $I^0_p = I(0)$. These current nodes have expressions of 
\begin{align}
    I_p^i &= I_p^{i-1} + u^{i-1}\Delta \alpha^i,  \quad \forall i \in 1..k.   \\
    \intertext{The signal energies contained within each interval are}
         E^i_{p} &= \begin{cases}
        (I_p^{i-1})^2 (\Delta \alpha^i) & u^i = 0 \\
        ((I_p^i)^3 - (I_p^{i-1})^3)/(3 u^i) & u^i \neq 0 \end{cases} \\
        E^i_{\text{mix}} &= 
            \gamma_0(I_p^{i-1}(\sin(\alpha^i + \phi) - \sin(\alpha^{i-1} + \phi)) \\
            &\quad +  u^i (\cos(\alpha^i + \phi) -  \cos(\alpha^{i-1} + \phi)) \\
            & \quad +   u^i (\Delta \alpha^{i} \sin(\alpha^i + \phi) -  \sin(\alpha^{i-1} + \phi))).
\end{align}

The total energy of each contribution is 
\begin{align}
    E_p &= \textstyle\sum_{i=1}^k E^i_p & E_\text{mix} &= \textstyle\sum_{i=1}^k E^i_\text{mix}. \label{eq:energy_sum}
\end{align}

\subsection{Mixed Load Characteristic}

We now consider the case of $\tau > 0$ with angle $\psi = \tan^{-1}(\tau)$. The current dynamics of \eqref{eq:dynamics_per_unit} are $\forall i \in 1..k \nonumber:$
\begin{align}
    I_p^i &= (u^{i-1}/\tau) + (I_p^{i-1} - u^{i-1}/\tau) \exp(- \tau \Delta \alpha^i). 
    \end{align}

    The energy contained purely in the pulse pattern signal is    
    \begin{subequations}    
    \begin{align}
    E^i_{p0} &= (u^i-\tau I^{i-1}_p)(3 u^i + \tau I^{i-1}_p) \\
    E^i_{p\Delta 1} &= 2 (u^i)^2 \tau \Delta \alpha^i\\
    E^i_{p\Delta 2} &= \exp(-2 \tau \Delta \alpha^i) (u^i - \tau I^{i-1}_p)\\
    & \qquad \qquad (\tau I^{i-1}_p + u^i(4 \exp(\tau \Delta \alpha^i) - 1))\nonumber\\
    E^i_{p} &= \left( E^i_{p1} +  E^i_{p2} - E^i_{p0}\right)/(2 \tau^3).
    \end{align}
\end{subequations}

    The energy contained in the mixture $I_p I_{\text{ext}}$ is
    \begin{subequations}
    \begin{align}
    \gamma &=  \gamma_0/(\tau^2+1) \\
    E^i_{s} &= (I^{i-1}_p \tau + u^{i-1}((\tau^2 +1)\exp(\tau \Delta \alpha^i)-1)  \\
    & \qquad \qquad  \sin(\alpha^i + \psi + \phi) \\
    E^i_{s0} &= (I^{i-1}_p \tau + u^{i-1}(\tau^2) \sin(\alpha^{i-1} + \psi + \phi) \\
    E^i_{c} &= -\tau(I^{i-1}_p \tau - u^{i-1}) \cos(\alpha^i + \psi + \phi) \\
    E^i_{c0} &= -\tau(I^{i-1}_p \tau - u^{i-1}) \cos(\alpha^{i-1} + \psi + \phi) \\
    E^i_{\text{mix}} &= \gamma(\exp(-\tau \Delta \alpha^i)(E^i_{s} + E^i_{c})- E^i_{s0} -E^i_{c0}).
\end{align}    
\end{subequations}

The energy of each component is summed as in \eqref{eq:energy_sum}
\begin{align}
    E_p &= \textstyle\sum_{i=1}^k E^i_p & E_\text{mix} &= \textstyle\sum_{i=1}^k E^i_\text{mix}. 
\end{align}

%% file: appendix/app_symmetry.tex
\section{Symmetry and Unipolarity}
\label{app:sym_unipolar}

This section discusses how \ac{HW} and \ac{QW} symmetries with possible unipolarity constraints can be incorporated into the hybrid optimal control problem \ref{prob:tdd_hy}.

\subsection{Transition Graph}
\begin{figure}[t!]
    \centering
    \includegraphics[width=0.9\linewidth]{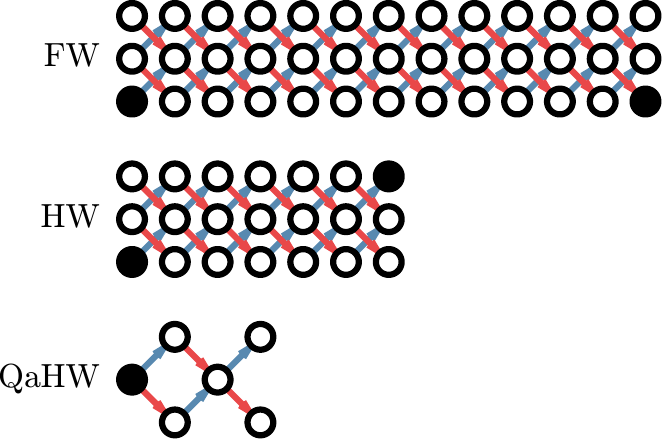}
    \caption{Transition graphs under symmetry}
    \label{fig:transition_symmetry}
\end{figure}

Symmetries and unipolar constraints can be imposed by restricting the structure of the transition graph $\gs$. Figure \ref{fig:transition_symmetry} visualizes transition graphs for the cases of \ac{FW} (top), \ac{HW} (middle) and \ac{QW} (bottom) symmetry. The black circles mark cases a  possible initial level for a periodic path. Only the vertices from $i \in 0..k/2$ ($k/4$) need to be considered for the \ac{HW}  (\ac{QW}) case: the rest will be filled in by symmetry. The \ac{HW} pattern furthermore requires that $n^{k/2} = N - n^{0}$ in order to ensure that $n^0 = n^{k}$. \ac{QW} structure requires that that $N$ is odd and $n^{0} = (N+1)/2$. Due to this restriction, vertices with $\abs{n+i}$ odd can are unreachable, and can be omitted from the description in $\gs$.

Unipolarity in the \ac{HW} or \ac{QW} settings can be enforced by removing all vertices with $n < (N+1)/2$ from the transition graph $\gs$ over the first half-period. Figure \ref{fig:transition_symmetry} draws a unipolar-constrained graph $\gs$ in the case of $N=7$ and $k=12$ for a \ac{QW}-constrained pulse pattern.

\begin{figure}[t!]
    \centering
    \includegraphics[width=0.9\linewidth]{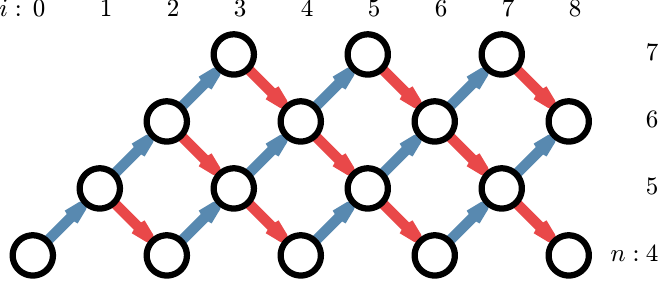}
    \caption{Transition graphs under unipolarity}
    \label{fig:transition_unipolarity}
\end{figure}

\subsection{Support Sets}

There are two possible approaches to consider when treating symmetry in the Hybrid OCP setting. The first option is to take the full $2\pi$ period. The second option is to track only the first $\pi$ (HW) or $\pi/2$ (QW) parts of the fundamental period, and then complete the signal by symmetry. If the first approach is taken, then the support sets from Section \ref{sec:support_set} can be kept the same. In the case of the second approach, symmetry-modified restricted support sets can be defined.

The specialized \ac{QW} constructions developed in this subsection are only valid in the case of $\tau = 0$ and $\phi = 0$. If these conditions are not satisfied then the \ac{HW} construction support set should instead be used, after extending and constraining the QW transition graph $\gs$ to an HW-compatible graph.

The symmetry-accounted initial sets are
\begin{align}
    \text{HW}: & & X^{0}_{n, 0} &= \{1\} \times \{0\} \times  [0, \pi -\Theta k/2] \times I_S  \label{eq:set_initial_qw} \\
    \text{QW}: & & X^{0}_{n, 0} &= \{1\} \times \{0\} \times  [\Theta/2, \pi/2 -\Theta k/4] \times I_S.  \nonumber
\end{align}

The clock range in the symmetry-modified formulation is smaller than in the full expression \eqref{eq:set_initial} of $[0, 2\pi - \Theta k]$. Note that the \ac{QW} initial set requires $\phi \geq \Theta/2$ due to the reflection about the angle $\theta = 0$.

The support sets modified from \eqref{eq:set_flow} are 
\begin{align}
    \text{HW}:& & X_{n, i} &= B([0, \pi]) \times [0,  \pi - \Theta(k/2)] \times I_S \\
    \text{QW}:& & X_{n, i} &= B([0, \pi/2]) \times [0,  \pi/2 - \Theta(k/4)] \times I_S. \nonumber
\end{align}

The terminal sets in the \ac{HW} or \ac{QW} setting are not the same as the initial set in \eqref{eq:set_initial}. The symmetry-modified terminal sets  are.
\begin{align}
    \label{eq:support_mode_terminal_sym}
   \text{HW}:& &  X_{n, d}^{T} &=  \{-1\} \times \{0\} \times  [\Theta/2, \pi- \Theta (k/2)] \times I_s \nonumber\\
   \text{QW}:& & X_{n, d}^{T} &=  \{0\} \times \{1\} \times  [\Theta/2, \pi/2 - \Theta (k/4)] \times I_s 
\end{align}
The HW (QW) system thus tracks the angle for $\theta \in [0, \pi]$ ($[0, \pi/2]$).

If a $k$-length \ac{HW} pulse pattern starts at a mode $(n, 0)$, then it must terminate at mode $(N-n, k/2)$  in the HW-restricted transition graph. A $k$-length pulse pattern for a a QW-symmetric pulse pattern must start at mode $((N+1)/2, 0)$, and can end at any mode $(n, k/4)$ such that $(n+k/4)$ is even.

\subsection{Integration}

We now develop integration maps in the case of HW ($\theta \in [0, \pi]$) and QW ($\theta \in [0, \pi/2]$) symmetries. 
We define the matrices
\begin{align}
    \Gamma_{\text{HW}} &= \begin{bmatrix}
        -1 & 0 & 0 & 0\\
        0 & -1 & 0 & 0\\
        0 & 0 & 1 & 0\\
        0 & 0 & 0 & -1\\
    \end{bmatrix} &\Gamma_{\text{HW}} &= \begin{bmatrix}
        0 & 1 & 0 & 0\\
        1 & 0 & 0 & 0\\
        0 & 0 & 1 & 0\\
        0 & 0 & 0 & 1\\
    \end{bmatrix}
\end{align}
and the orbit operators (weighted group averages)
\begin{align}
    \mathcal{R}_{\text{HW}}[z(x, u)] &= z(x, u) + z(\Gamma_{\text{HW}}x, -u) \\
    \mathcal{R}_{\text{QW}}[z(x, u)] &= z(x, u) + z(\Gamma_{\text{QW}} x, u) \\
    & \qquad + z(\Gamma_{\text{HW}} x, -u) + z(\Gamma_{\text{HW}} \Gamma_{\text{QW}} x, -u). \nonumber
\end{align}

HW integrators can be defined for functions $z(x, u)$ as 
\begin{subequations}
    \label{eq:integral_operator_sym}
\begin{align}
    \Lambda_\vs^\mathcal{T}: z(x, u) &\mapsto  \sum_{(n, i) \in \vs} \int_{\theta=0}^{\pi} \mathcal{R}_{\text{HW}}\chi_{(n, i)}^\mathcal{T}(\theta)z(x(\theta), u_n)] d \theta \\ 
        \Lambda_\es^\mathcal{T}: z(x, u) &\mapsto  \sum_{e \in \es} 
        \mathcal{R}_{\text{HW}}[\chi_{(n, i)}^\mathcal{T}(\theta)z(x(\theta), u_n)] d \theta.
\end{align}
\end{subequations}
QW integrators have a similar formula, but with limits $[0, \pi/2]$ and averaging operators $\mathcal{R}_{\text{HW}}$.
Simplified formulae exist for \eqref{eq:integral_operator_sym} if the the function $z$ matches the symmetry structure of the pattern. If $z$ is HW-symmetryic, the HW mode-integral  operator from \eqref{eq:integral_operator_sym} has an expression 
\begin{align}
    \Lambda_\vs^\mathcal{T}: z(x, u) &\mapsto  2 \sum_{(n, i) \in \vs} \int_{\theta=0}^{\pi} \chi_{(n, i)}^\mathcal{T}(\theta)z(x(\theta), u_n) d \theta.
\end{align}
The averaging $\mathcal{R}_{\text{HW}}$ is replaced by a factor of 2 in front of a sum. If $z$ is QW symmetric or QW anti-symmetric, the average operator $\mathcal{R}_{\text{QW}}$ is replaced by a factor of 4.

%% file: appendix/app_measure_theory.tex
\section{Measure Theory and Occupation Measures}
\label{app:measure}
This appendix summarizes mathematical concepts in measure theory and (hybrid) occupation measures for dynamical systems. The measure constructions will be used to construct the convex relaxations to Problem \ref{prob:tdd_hy}.

\subsection{Analysis}
\label{app:analysis}

We first define notation and review measure-theoretic concepts.
The set of continuous functions over a set $S$ is $C(S)$. Its subring of nonnegative functions over $S$ is $C_+(S)$. The set of $k$-times differentiable functions over $S$ is $C^k(S).$  The set of signed Borel measures supported over $S$ is $\mathcal{M}(S)$, and its subset of nonnegative Borel measures is $\Mp{X}$. When the set $S$ is compact, the sets $C_+(S)$ and $\Mp{S}$ possess an inner product $\inp{\cdot}{\cdot}$ by Lebesgue integration: $\forall f_+ \in C_+(S), \ \mu \in \Mp{S}: \inp{f_+}{\mu} = \inp{f_+(s)}{\mu(s)} = \int_S f(s) d \mu(s)$. This Lebesgue-integration-based inner product  $\inp{\cdot}{\cdot}$ can be lifted into a duality pairing between $C(S)$ and $\Mp{S}$ as $\forall f_+ \in C(X), \ \mu \in \Mp{S}: \inp{f}{\mu} = \int_S f(s) d \mu(s)$.

Given two measures $\mu \in \Mp{S_1}, \nu \in \Mp{S_2}$, the product measure $\mu \otimes \nu$ is defined as $\forall f_1 \in C(S_1), f_2 \in C(S_2): \ \inp{f_1(s_1) f_2(s_2)}{\mu_1(s_1) \mu_2(s_2)} = \inp{f_1}{\mu_1}\inp{f_2}{\mu_2}$. A map $R: S_1 \mapsto S_2$ induces a pushforward of measures $\mu \in \Mp{S_2}$ as $R_\# \mu$ with the relation $\forall f \in C(S_2), \ \mu \in \Mp{S_1}: \inp{f(R(s_1))}{\mu(s_1)} = \inp{f(s_2)}{R_\# \mu(s_2)}$. The projection map $\pi^{s_1}$ acts on a tuple $(s_1, s_2)$ as $\pi^{s_1}(s_1, s_2) = s_1$ (with a similar action for $\pi^{s_2}$). The pushforward of the projection map $\pi^{s_1}_\#$ therefore serves as the $s_1$-marginalization operator ($\forall f \in C(S_1), \eta \in \Mp{S_1 \times S_2}: \ \inp{f(s_1)}{\eta(s_1, s_2)} = \inp{f(\pi^{s_1}(s_1, s_2))}{\eta(s_1, s_2)} =  \inp{f(s_1)}{\pi^{s_1}_\#\eta(s_1)}$.

The mass of a measure $\mu$ is $\inp{1}{\mu}.$ The measure $\mu$ is a probability measure if $\inp{1}{\mu} = 1$.

The set of polynomials with real coefficients in an indeterminate variable $s$ is $\R[s]$.
The set of $n$-dimensional multi-indices is $\N^n$, and the exponential notation $s^\eta$ will be used to abbreviate the monomial $\prod_{i=1}^n s_i^{\eta_i}$.
Given an exponent $\eta \in \N^n$,  the $\eta$-moment of a measure $\mu(s)$ is $m_{\eta} = \inp{s^\eta}{\mu(s)}$. The degree of a polynomial $p \in \R[s]$ is $\deg p$, and the set of polynomials of degree $\leq \beta$ is $\R[s]_{\leq \beta}$


\subsection{Standard and Occupation Measures}

\label{sec:occ_measures}
The Dirac delta $\delta_{s'} \in \Mp{S}$ is a probability measure supported only at the point $s'$, following the pairing $\forall f \in C(S): \ \inp{f}{\delta_s'} = f(s').$ Given a set  $T \subseteq S$, the Lebesgue measure $\lambda_T$ is the measure satisfying $\forall f \in C(S): \ \inp{f}{\lambda_T} = \int_{T} f(s) ds$. The zero measure $\0 \in \Mp{S}$ satisfies the pairing $\forall f \in C(S): \inp{f}{0} = \0$.

Given a curve $x(t): [0, T] \rightarrow X$, the occupation measure $\mu_x$ of the curve is the measure satisfying $\forall A \subseteq [0, T], \ B \in X$
\begin{align}
    \mu_x(A \times B) = \int_{t=0}^T I_{A \times B}(t, x(t)) dt. \label{eq:occ_measure}
\intertext{Relation \eqref{eq:occ_measure} implies that for all test functions $w \in C^1([0, T] \times X)$ it holds that}
    \inp{w}{\mu_x} = \int_{t=0}^T w(t, x(t)) dt.
\end{align}
The occupation measure in \eqref{eq:occ_measure} is therefore the pushforward of $\lambda_{[0, T]}$ along the curve-graph map $t \mapsto (t, x(t))$. 

Given a function $f(t, x),$ we denote the Lie derivative $\Lie_f$ as the map $v(t, x) \mapsto \partial_t v(t, x) + f\cdot \nabla_x v(t, x)$.
If $x(t)$ arises from the solution of an ordinary differential equation $\dot{x}(t) = f(t, x)$, then the evaluation of any arbitrary function $v \in C^1([0, T] \times X)$ satisfies a conservation law $\forall 0 \leq t \leq t' \leq T$
\begin{align}
    v(T, x(T)) = v(0, x(0)) + \int_{t=0}^T \Lie_f v(t, x(t)) dt. \label{eq:conservation}
\end{align}
The conservation law in \eqref{eq:conservation} may be expressed in using the occupation measure $\mu_x$
\begin{align}
    \inp{v}{\delta_{T, x(T)}} = \inp{v}{\delta_{0, x(0)}} + \inp{\Lie_f v}{\mu_x}.
\end{align}
This conservation law is a specific instance of a Liouville equation, and it may be further generalized to initial measures $\mu_0$, terminal measures $\mu_T$, and occupation measure $\mu$ satisfying 
\begin{align}
    \inp{v}{\mu_T} = \inp{v}{\mu_0} + \inp{\Lie_f v}{\mu}. \label{eq:liouville}
\end{align}

When $f$ is a Lipschitz function and $X$ is compact, any tuple ($\mu_0, \mu_T, \mu$) satisfying \eqref{eq:liouville} is supported on the graph of a (possibly infinite) convex combination of trajectories of $\dot{x} = f(t, x)$ (Theorem 3.1 of \cite{lewis1980relaxation}).

\subsection{Linear Programs in Measures}

Convex optimization problems may be posed over infinite-dimensional quantities such as measures \cite{barvinok2002convex, fattorini1999infinite}.

A linear program over $s$ measures $\forall i \in 1..s: \ \mu_i \in \Mp{\mathcal{X}_i}$ has the form of 
\begin{align}
    p^* &= \inf_{\mu} \ \textstyle \sum_{i=1}^s \inp{c_i}{\mu_i} & \mathcal{A}(\mu) = b \label{eq:lin_prog_meas}
\end{align}
with respect to a cost functions $c_i \in C(\mathcal{X}_i),$ an answer vector $b \in \mathcal{Y}$, and an affine map $\mathcal{A}: \prod_{i=1}^s \Mp{\mathcal{X}_i} \rightarrow \mathcal{Y}$. A topological dual linear program to \eqref{eq:lin_prog_meas}  $v \in \mathcal{Y}'$ is
\begin{align}
    d^* &= \sup_{v} \ \inp{v}{b} & \mathcal{A}^*(v)  + c\geq 0. \label{eq:lin_prog_func}
\end{align}
If $(\mathcal{A}, b, c)$ are continuous, feasible solutions $\mu$ to $\mathcal{A}(\mu) =b$ are bounded, a feasible  $\mu$ exists, and each $\mathcal{X}_i$ is compact, then strong duality will occur $p^* = d^*$ with attainment of optima \cite[Theorem 2.6]{tacchi2021thesis}. 

\subsection{Moment-Based Truncations}

The primal-dual pair in \eqref{eq:lin_prog_meas} and \eqref{eq:lin_prog_func} are infinite-dimensional convex optimization problems. The moment-SOS hierarchy \cite{lasserre2009moments} offers one method to bound the infinite-dimensional linear programs by an outer-approximating sequence of finite-dimensional semidefinite programs in the case where $(\mathcal{A}, b, c)$ are all described by polynomials. For further detail on all preliminary content in this subsection, refer to \cite{lasserre2009moments, henrion2020moment}.
A measure $\mu \in \Mp{\mathcal{X}}$ with $\mathcal{X} \subseteq \R^{n}$ has an associated infinite-dimensional moment sequence $\{m_{\eta}\}_{\eta \in \N^n} = \{\inp{s^\eta}{\mu(s)}\}_{\eta}$. To any $\eta$-indexed set of numbers $\{y_\eta\}$, a linear (Riesz) functional  $\mathbb{L}[y]: \R[x] \rightarrow \R$ exists under the definition $    \mathbb{L}[y](\sum_{\eta} c_\eta x^\eta) \rightarrow \sum_{\eta} c_\eta y_\eta.$  In the case where the sequence $y$ is the moment sequence of a measure $(y = m)$, then for any polynomial $p \in \R[x]$ it holds that $\mathbb{L}[m](p) = \inp{p}{\mu}$.

Nonnegative Borel measures satisfy nonnegativity properties: for any real-valued $g \in C_+(\mathcal{X})$ and $q \in C(X)$, it holds that $\forall x \in \mathcal{X}: g(x) q(x)^2 \geq 0$ and $\inp{g q^2}{\mu} \geq 0$. In the case where $\mathcal{X}$ is described by a locus of nonnegativity of $N_c$ inequality constraints and $N_i$ inequality constraints with $g_0(x) = 1$ as
\begin{align}
    \mathcal{X} = \{ g_i(x) \geq 0, h_j(x) = 0 & & \forall i \in 0..N_c, \ j \in 1..N_i\}, \label{eq:X_set}
\end{align}
with $N_c, N_i$ finite, the measure $\mu$ satisfies the following nonnegativity relations $\forall q \in C(X)$:
\begin{align}
    \forall i: & \ \inp{g_i q^2}{ \mu} \geq 0 & \forall j: &  \ \inp{h_j q^2}{\mu} = 0. \label{eq:nonneg_meas}
\end{align}
The relation \eqref{eq:nonneg_meas} induces constraints in any sequence $\{y_\eta\}$:
\begin{align}
    \forall i: & \ \mathbb{L}[y](g_i q^2) \geq 0 & \forall j: &  \ \mathbb{L}[y](h_j q^2)  = 0. \label{eq:nonneg_riesz}
\end{align}

When the constraint-describing functions $g_i, h_j$ from \eqref{eq:X_set} are all polynomial in $x$ ($g_i = \sum_\eta c^i_\eta x^\eta, \ h_j = \sum_{\eta} \tilde{c}^j_\eta x^\eta$), the set $\mathcal{X}$ is known as a Basic Semialgebraic (BSA) Set. Imposition of \eqref{eq:nonneg_riesz} with respect to all $q \in \R[x]$ (loosened from $q \in C(X)$) results in a set of convex constraints in the elements $y$.  
A Localizing matrix may be defined for each $g_i$: 
\begin{align}
    \mathbb{M}[g_i y]_{\eta_1, \eta_2} = \textstyle \sum_{\gamma \in \R^n} c^i_{\gamma}y_{\gamma + \eta_1 + \eta_2}.
\end{align}
Given a finite degree $\beta \in \N$, we denote $\mathbb{M}_{\beta- \lceil \deg g_i/2 \rceil}[g_i y]$ as the finite-dimensional square top-left-corner submatrix of $\mathbb{M}[g_i y]$ containing elements $y$ only of degree $\leq 2\beta$. The size of this submatrix $\mathbb{M}[g_i y]$  $\binom{n+\beta - \lceil \deg g_i/2 \rceil}{n}$. The degree-$\beta$ restriction of relation \eqref{eq:nonneg_riesz} is 
\begin{align}
    \forall i: & \ \mathbb{M}_{\beta- \lceil \deg g_i/2 \rceil}[g_i y] \succeq 0 & \textstyle \forall j: &  \ \sum_\gamma \tilde{c}^j_\gamma y_{\eta + \gamma} = 0. \label{eq:nonneg_restr}
\end{align}
We denote $\mathbb{M}_{\beta}[\mathcal{X}_i y] \succeq 0$ as the collection of restrictions in \eqref{eq:nonneg_restr} to simplify notation.

Algebraic reductions (grobner bases) can be used to simplify the equality constraints and reduce the size of the Localizing matrices by restriction to a quotient ring \cite{parrilo2005exploiting}. As an example (used in this work), a measure supported on $B = \{(c, s) \mid c^2 + s^2 = 1\}$ satisfies $ \forall \eta \in \N^2: \ m_{(2, 0)+\eta} = m_{(0, 0) + \eta} - m_{(0, 2) + \eta}$. As such, any term $y_\gamma$ with $\gamma \geq (2, 0)$ may be replaced by other entries of $y$. A pseudomoment sequence $y$ posed over $n$ state variables with maximal degree $2\beta$ in which two of the variables are restricted to $B$ (a single quadratic constraint) involves a total of $\binom{n-1+2\beta}{2\beta} + \binom{n+2\beta - 1}{2\beta-1} < \binom{n + 2\beta}{2\beta}$ parameters in $y$ restricted by PSD constraints of maximal size $\binom{n-1 + \beta}{\beta} + \binom{n+\beta -1}{\beta-1} < \binom{n+\beta}{\beta}$ due to this algebraic simplification. We denote $\N^{\circ n}_{\beta} \subset \N^n$ as the set of $\binom{n-1 + \beta}{\beta} + \binom{n+\beta -1}{\beta-1}$ multi-indices satisfying for $\gamma \in \N^{n}$ the property $\gamma_1 < 2$ and $\sum_{i=1}^n \gamma_i \leq 2\beta$.


The moment side of the moment-SOS hierarchy involves replacement of the measure formulation in \eqref{eq:lin_prog_meas} to the satisfaction of linear matrix inequality constraints:
\begin{subequations}
\begin{align}
    p^*_\beta = &\inf_{y} \ \textstyle \sum_{i=1}^s \mathbb{L}[y^i](c_i) \\
    & \mathbb{A}_\beta (y) = b , \  \forall i: \mathbb{M}_\beta[\mathcal{X}_i y_i ] \succeq 0.
\end{align}
\label{eq:lin_prog_mom}
\end{subequations}

The operator $\mathbb{A}_\beta$ is the restriction of the linear operator $\mathcal{A}$ to the set of measures degree-2$\beta$ pseudo-moment sequences $y$. The time complexity of solving \eqref{eq:lin_prog_mom} at fixed $\beta$ scales as $O(s^{3/2}\beta^{9/2n})$ in the case that all measures have the same number of states \cite[Section 3.4]{claeys2016modal}.  The set of optima will rise as $p^*_{1} \leq p^*_{2} \leq \ldots p^*_{\beta} \ldots \leq p^*$, given that increasing the degree $\beta$ tightens the set of constraints. This sequence will converge as $\lim_{\beta \rightarrow \infty} p^*_\beta = p^*$ if the sets $\mathcal{X}_i$ satisfy a ball (Archimedean/compactness) constraint: that there exists a $R > 0$ such that the polynomial $R - \norm{x}_2^2$ is inside the quadratic module of the constraint-describing polynomials $(g, h)$ for each $\mathcal{X}_i$. 
This ball constraint holds whenever $\mathcal{X}_i$ is a ball, a box, or (relevant to this work) the intersection between a cylinder $(c, s)$ and a box $(\phi, I)$. \label{app:mom_trunc}

%% file: appendix/app_convex_relaxation.tex
\section{OPP Convex Relaxation}
\label{app:measure_program}

This appendix presents the convex relaxation used to lower-bound the TDD in OPPs. The optimal control problem in Problem \ref{prob:tdd_hy} is first reformulated as a linear program in measures \eqref{eq:lin_prog_meas}, and then truncated through the moment-SOS hierarchy into a sequence of semidefinite programs \eqref{eq:lin_prog_mom}.

\subsection{Measure Variables}

Table \ref{tab:meas_tdd} details the measure variables used in the measure LP representation of Problem~\ref{prob:tdd_hy}.

\begin{table}[t!]
    \centering
    \caption{Measure variables used in OPP program}
    \begin{tabular}{l r  l l}
        Initial & $\mu^0_n$ & \hspace{-0.3cm}$\in \Mp{X^0_n}$ &  n \\
        Terminal & $\mu^T_{n}$ &  \hspace{-0.3cm}$\in \Mp{X^{T}_{n}}$ & $n$ \\
        Occupation & $\mu_{n, i} $ &  \hspace{-0.3cm}$\in \Mp{X_{n, i}}$ &  $(n, i) \in \vs$ \\
         Step Up & $\rho_{n, i}^+$ &  \hspace{-0.3cm}$\in \Mp{G_{n, i}^+}$ & \hspace{-0.2cm} $(n-1, i-1) \rightarrow (n, i)\in \es^+$ \\
         Step Down & $\rho_{n, i}^-$ &  \hspace{-0.3cm}$\in\Mp{G_{n, i}^-}$& \hspace{-0.2cm} $(n+1, i-1) \rightarrow (n, i)\in \es^-$ \\
    \end{tabular}
    \label{tab:meas_tdd}
    \vspace{-8pt}
\end{table}

The initial measure $\mu^0$ tracks the clock $\phi$ and current $I$ before any switching is performed. The terminal measure $\mu^T$ tracks $\phi$ and $I$ at the end of the trajectory (implicit for FW and HW, left free for QaHW). The occupation measure $\mu_{n, i}$ stores trajectory information between switches number $i$ and  $i+1$ when following linear dynamics \eqref{eq:dynamics}. The measures $\rho^\pm$ store the switching angle, clock angle, and current at which point the level $u$ increases or decreases.

\subsection{Construction Procedure}
\label{sec:construction}
Measures from Table \ref{tab:meas_tdd} may be constructed from a feasible pulse pattern  $\mathcal{T}$. As an example, these constructed measures will satisfy the integration relation for all test functions $z(x, u)$:
\begin{align}
    \Lambda_\vs^\mathcal{T}[z] = \sum_{(n, i) \in \vs} \inp{z(x, u_{n^i})}{\mu(c, s, \phi, I)}.\label{eq:integral_relation}
\end{align}

Let $I(\theta)$ denote the load current when the input $u(\theta)$ derived from pulse pattern $\mathcal{T}$ is applied. 
The initial measure derived from the pattern $\mathcal{T}$ is
\begin{align}
    \mu_{n, i}^0 = \begin{cases}
        \delta_{\phi = 2\pi - \alpha^k, \ I = I(0)} & n = n^0 \\
        0 & \text{else}
    \end{cases}.
\end{align}
The terminal measure (with respect to symmetry) is 
\begin{subequations}
\begin{align}
    \text{FW} & & \mu_{n, i}^T &= \mu^0_{n, i} \\
    \text{HW} & & \mu_{n, i}^T &= (\phi, -I)_\# \mu^0_{n, i} \\
    \text{QaHW} & & \mu_{n, i}^T &= \begin{cases}
        \delta_{\phi = \pi/2 - \alpha^{k/2}, \ I = I(\pi/2)} & n = n^{k/2} \\
        0 & \text{else}.
    \end{cases}
\end{align}
\end{subequations}

The occupation measures selected to satisfy $\forall w \in C(X)$ the following relation:
\begin{align}
   \textstyle \inp{w}{\mu_{n, i}} = \int w(x) d \tilde{\mu}_{n, i}(x) = \int_{\theta=0}^{2\pi} \chi_{n, i}^\mathcal{T}(\theta) w(x(\theta)) d \theta. \label{eq:occ_int}
\end{align}
The occupation measures have upper limits $\pi$ and $\pi/2$ for  HW and QaHW respectively, instead of $2\pi$ from FW.
The switching measures are defined as
\begin{align}
    \tilde{\rho}^\pm_{n, i} = \begin{cases}
        \delta_{(c, s) = \psi(\alpha^i), \phi = \alpha^i - \alpha^{i-1}, I = I(\alpha^i)} & n = \text{dst}(P^i), P^i \in \mathcal{E}^\pm \\
        0 & \text{else}.
    \end{cases}
\end{align}

If the pattern $\mathcal{T}$ is feasible for the constraints of Problem \ref{prob:tdd_hy}, then the measures constructed in this subsection will be feasible for the measure LP defined in the subsequent subsection.

\subsection{Measure Program Terms}

Table \ref{tab:meas_con} summarizes the terms used to describe the OPP measure program. 

\begin{table}[t!]
    \centering
        \caption{Expressions in OPP Linear Program}
    \begin{tabular}{lc}
        Term &  Measures  \\ \hline
        Objective &  $\mu$ \\
        Probability & $\mu^0$ \\
        Harmonics & $\mu$ \\
        Continuity & All \\
        Uniformity & $\mu$    \\
        Quarter-Matching & $\rho^\pm$
    \end{tabular}
    \label{tab:meas_con}
\end{table}

Expressions may involve the symmetry factor $C_{\text{sym}}$
\begin{align}
C_{\text{sym}} =     \begin{cases}
        1 & \text{FW Symmetry} \\
        2 & \text{HW Symmetry} \\
        2 & \text{QW Symmetry, $\tau \in (0, \infty)$} \\
        4 & \text{QW Symmetry, $R_\text{load}=0$ or $L_\text{load}=0$.}        
    \end{cases}
\end{align}
The number of switching transitions analyzed in an OPP setup is therefore $k' = k/C_{\text{sym}}$.
We now proceed to explain each expression.
\subsubsection{Objective}

The objective  in \eqref{eq:tdd_hy_obj} is the signal energy $\Lambda^{\mathcal{T}}_\vs[I^2]$. This objective has the formulation
\begin{align}
    \Lambda^{\mathcal{T}}_\vs[I^2] &= C_{\text{sym}} \textstyle \sum_{v \in \vs} \mu_{v}.
\end{align}

\subsubsection{Probability}

The initial measure must be normalized to be a probability distribution, in order to include only a single curve as an admissible solution. This constraint is
\begin{align}
    \textstyle \sum_{n \in L} \inp{1}{\mu^0_{n, 0}} = 1. \label{eq:con_probability}
\end{align}

The QaHW case only includes the central level $(N+1)/2$ as an admissible start point. As such, the probability constraint in this case is supplemented to  $\inp{1}{\mu^0_{(N+1)/2, 0}} = 1,$ which enforces $\mu^0_{n, i} = 0$ if $n \neq (N+1)/2$.

\subsubsection{Harmonics}

The harmonics specifications can be enforced through affine constraints involving Chebyshev polynomials
\begin{subequations}
    \label{eq:con_harmonic}
\begin{align}
   \textstyle  C_{\text{sym}} \sum_{(n, i) \in \vs} u_{n^i} \inp{s U_{q^b}(c)}{\mu_{(n, i)}} \in h^b.
\intertext{In the case of FW symmetry, the cosine constraints can be enforced by}
    \textstyle \sum_{(n, i) \in \vs} u_{n^i} \inp{T_{q^a}(c)}{\mu_{(n, i)}} \in h^a.
\end{align}
\end{subequations}
In these definitions, it is assumed that harmonics indices $q^a$ and $q^b$ are not present if they are identically zero by HW or QaHW symmetry.

\subsubsection{Continuity}

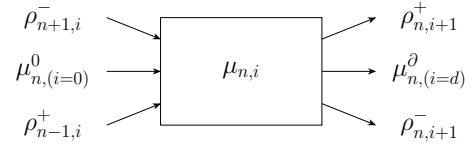
\begin{figure}[t!]
\centering
\resizebox{0.7\linewidth}{!}{%
\begin{circuitikz}
\tikzstyle{every node}=[font=\LARGE]
\draw  (7.5,13) rectangle  node {\LARGE $\mu_{n, i}$} (11.25,10.5);
\draw [->, >=Stealth] (6.25,11.75) -- (7.5,11.75);
\draw [->, >=Stealth] (11.25,11.75) -- (12.5,11.75);
\draw [->, >=Stealth] (11.25,12.5) -- (12.5,13);
\draw [->, >=Stealth] (11.25,11) -- (12.5,10.5);
\draw [->, >=Stealth] (6.25,10.5) -- (7.5,11);
\draw [->, >=Stealth] (6.25,13) -- (7.5,12.5);
\node [font=\LARGE] at (5,11.75) {$\mu_{n, (i=0)}^0$};
\node [font=\LARGE] at (5,10.5) {$\rho_{n-1, i}^+$};
\node [font=\LARGE] at (5,13) {$\rho_{n+1, i}^-$};

\node [font=\LARGE] at (13.75,11.75) {$\mu_{n, (i=d)}^\partial$};
\node [font=\LARGE] at (13.75,10.5) {$\rho_{n, i+1}^-$};
\node [font=\LARGE] at (13.75,13) {$\rho_{n, i+1}^+$};
\end{circuitikz}
}%
\caption{Continuity relation between measures}
\label{fig:continuity}
    \vspace{-6pt}
\end{figure}

Measures from Table \ref{tab:meas_tdd} are related by the dynamical laws \eqref{eq:mode_dynamics} and \eqref{eq:jump_dynamics}. This relation may be imposed through the enforcement of per-mode Liouville equation \eqref{eq:liouville} as covered in Appendix \eqref{sec:occ_measures}. The continuity relation between measures is visualized in Figure \ref{fig:continuity}. In particular, probability mass enters mode $(n, i)$ through a step up from $(n-1, i-1)$, a step down from $(n+1, i-1)$, or from the initial condition if $i=0$ at angle $\theta = 0$. Mass leaves mode $(n, i)$ through a step up to $(n+1, i+1)$, a step down to $(n-1, i+1)$, or through termination at $i = k/C_{\text{sym}}$ and angle $\theta = 2\pi/C_{\text{sym}}$.

Given a test function $w \in C^1(X)$ and an angle $\alpha$, we perform the following abbreviations:
\begin{align}
    w_\alpha(\phi, I) &= w(\cos(\alpha), \sin(\alpha), \phi, I)  \\
    w_\phi(c, s, I) &= w(c, s, 0, I).    
\end{align}
In particular $w_0(\phi, I)$ is the evaluation at the initial condition $\theta=0, \ (c, s) = (1, 0)$. Reset dynamics in \eqref{eq:jump_dynamics} are implemented by the symbol $w_\phi$. Furthermore for any function $w \in C^1(X)$, we define the symbol $\Lie_{n, i}$ to denote the Lie derivative of the dynamics \eqref{eq:mode_dynamics}:
\begin{align}
    \Lie_{n, i} w(x) = (-s \partial_c + c \partial_s + \partial_\phi + (u_{n^i} - \tau I) \partial_I)w(x).
\end{align}

The conservation equation for all functions $w \in C^1(X)$ (with $\mu_{n, i}=0$ if $(n, i) \not\in \vs$ \textit{mutadis mutandis} for $\rho^\pm, \mu^{\partial}$) and holding over all $n \in 1..N$ is:
\begin{align}
    &\inp{w_0}{\mu^0_{n}} + \inp{\Lie_{n, 0}w}{\mu_{n, 0}} = \inp{w}{\rho^+_{n+1, 1} + \rho^{-}_{n-1, 1}} \nonumber\\   
    & \forall i \in 1..k/C_{\text{sym}}-1:\label{eq:con_conservation}  \\
    & \  \ \inp{w_\phi}{\rho^-_{n+1, i} + \rho^+_{n-1, i}} +  \inp{\Lie_{n, i} w}{\mu_{n, i}} \nonumber \\
    & \qquad = \inp{w}{\rho^+_{n+1, i+1} + \rho^{-}_{n-1, i+1}} \nonumber\\
    &\inp{w_\phi}{\rho^-_{n+1, k'-1} + \rho^+_{n-1, k'-1}} + \inp{\Lie_{n, k'} w}{\mu_{n, k'}} \nonumber \\
    &  \qquad = \inp{w_{2\pi/C_{\text{sym}}}}{\mu^T_{n}}.\nonumber
\end{align}

The expected value of any test function $w$ at the terminal point $\theta=2\pi / C_{\text{sym}}$ will be equal to the value of $w$ at the starting point $0$ plus the accumulated change in $w$ across all traversed modes and arcs \cite{zhao2017optimal}.

\subsubsection{Uniformity}

The uniformity constraint is a redundant expression that aids in producing sensible numerical solutions under polynomial optimization. This uniformity constraint is 
\begin{align}
    \textstyle \sum_{(n, i) \in \vs}  \pi^{c s}_\# \mu_{n, i} = \psi_\# \lambda_{[0, 2\pi/C_{\text{sym}}]}. \label{eq:con_uniformity}
\end{align}
The uniformity constraint ensures that the pulse pattern is entirely defined in the relevant angular arc $\theta \in [0, 2\pi/C_{\text{sym}}]$. The uniformity constraint is required because the function $\text{atan2}(c, s)$ is discontinuous and non-polynomial in $[0, 2\pi]$, and therefore cannot be applied as a test function in the continuity constraint (given that $\text{atan2} \not\in C^1(B)$).

\subsubsection{Quarter-Matching}
Quarter-matching constraints are only employed when QW symmetry is enforced, and are not used in the HW or FW settings.
Figure \ref{fig:symmetries} visualizes how a QW-symmetric applied input $u(\theta)$ will yield an HW-symmetric current output $I(\theta)$ when $\tau \in (0, \infty)$. As a result, the current $I(\theta)$ must be tracked in the entire range $[0, \pi]$, even when the input $u$ is uniquely defined given its value in $[0, \pi/2]$. 
Quarter-Matching constraints enforce quarter-wave symmetry of the applied input $u(\theta)$. The quarter-wave matching constraints have the form of $\forall w \in C(B)$, $(n, i)\rightarrow(n\pm 1, i+1) \in \es$:
\begin{align}
    &\inp{w(c, s)}{\rho^{\pm}_{(n, i) \rightarrow (n\pm 1, i+1)}(c, s, \phi, I)}   \label{eq:quarter_matching}\\
    \qquad &= \inp{w(-c, s)}{\rho^{\mp}_{(n, k/2-i) \rightarrow (n\mp 1, k/2-i+1)}(c, s, \phi, I)} \nonumber
\end{align}

As an example, if the edge $(2, 3) \rightarrow (3, 4)$ is traversed at $\theta = \pi/3$ for a QW-symmetric sequence with $k=16$ pulses, then a corresponding edge $(3, 5) \rightarrow (2, 6)$ must also be traversed in the transition graph at $\theta = 2\pi/3$.

\subsection{Measure Program}

The minimal-energy measure linear program  is
\begin{prob}
\label{prob:tdd_lp}
    Given device and design parameters in Table \ref{tab:device} and \ref{tab:design}, find measures $\mu$ to solve the following program
    \begin{subequations}        
    \label{eq:tdd_lp}
    \begin{align}
        p^* = &\inf_\mu \textstyle\ C_{\text{sym}} \sum_{n, i}  \inp{I^2}{\mu_{n, i}(x)} \\
        & \text{Constraints \eqref{eq:con_probability}, \eqref{eq:con_harmonic}, \eqref{eq:con_uniformity}, \eqref{eq:con_conservation}, \eqref{eq:quarter_matching}} \label{eq:con_all}\\
        & \text{Measures from Table \ref{tab:meas_tdd}.} 
    \end{align}
    \end{subequations}
\end{prob}

If it is assumed that the current $I$ is restricted to a compact set $I_S$ with all support sets in Table \ref{tab:meas_tdd} modified accordingly, then it will hold that $p^* = P^*$. This equivalence is because the per-mode dynamics are Lipschitz (linear) support sets are compact, and the reset map $(c, s, \phi, I) \rightarrow (c, s, 0, I)$ is continuous. Therefore, any solution to the conservation equation \eqref{eq:conservation} is supported over a superposition of trajectories of the per-mode (hybrid) linear dynamics (Lemma 8 of \cite{zhao2017optimal}). While the work in \cite{zhao2017optimal} requires stochastic kernels due to the presence of controllers within each mode (and thus needs differential inclusions), our setting involving decisions made over full-dimensional guards involves only per-mode ODE dynamics. 

Additionally, the considerations from \cite[Theorem 2.6]{tacchi2021thesis} hold, and thus the infimum will be attained by a minimum. The construction procedure from Section \ref{sec:construction} yields a feasible measure, the objective $I^2$ is bounded below by 0 and above by $\max_{I' \in I_s} (I')^2$, all measures are supported in compact sets, and the nonnegative mass of all nonnegative measures are bounded ($\sum \inp{1}{\mu^0_n}=1$ by Probability \eqref{eq:con_probability}, $\sum{\inp{1}{\mu_{n, i}}} = \frac{2\pi}{C_{\text{sym}}}$ by Uniformity \eqref{eq:con_uniformity}, $\sum \inp{1}{\mu^T_n}=1$ and $\sum \inp{1}{\rho^+_{n, i} + \rho^-_{n, i}}=1$ by Conservation \eqref{eq:conservation} with respect to the test function $w=1$).


\subsection{Moment Program}

We now describe the degree-$\beta$ truncation of Problem \eqref{eq:tdd_lp} via the moment-SOS hierarchy. Table \ref{tab:mom_tdd} describes the pseudomoment variables induced by the measures in Table \ref{tab:meas_tdd}.

\begin{table}[h]
    \centering
    \begin{tabular}{cccc}
    Sequence & States & Multiplicity & Max PSD. Size \\\hline
        $y^0_{n}$ & $\phi, I$ &  $L$ & $(1+\beta) + \binom{1+\beta}{2}$\\
         $y^T_n$ & $\phi, I$ & $L$  & $(1+\beta) + \binom{1+\beta}{2}$\\
         $y_{n, i}$ & $c, s, \phi, I$ &  $\abs{\vs}$ & $\binom{3 + \beta}{3} + \binom{3+\beta}{4}$ \\
         $y^\pm_{n, i}$ & $c, s, \phi, I$  & $\abs{\es}$ & $\binom{3 + \beta}{3} + \binom{3+\beta}{4}$
    \end{tabular}
    \caption{Variables of the OPP moment relaxation}
    \label{tab:mom_tdd}
\end{table}

The degree-$\beta$ moment relaxation to Program \eqref{eq:tdd_lp} is:
\begin{prob}
    Given a degree $\beta \geq 1$, and device and design parameters in Table \ref{tab:device} and \ref{tab:design}, find pseudomoment sequences $y$ to solve the following program:
    \label{prob:tdd_mom}
\begin{subequations}
\label{eq:tdd_mom}
\begin{align}
    p^*_{\beta} = &\min_y \textstyle C_{\text{sym}}\sum_{(n, i) \in \vs}   \mathbb{L}[y_{n, i}](I^2)\\
    & \M_\beta[X^0_n y_n^0], \ \M_\beta[X^T_n y_n^T] \succeq 0 \nonumber\\
    & \M_\beta[X_{n,i} y_n^T], \ \M_\beta[G^\pm_{n,i} \rho_{n,i}^\pm] \succeq 0 \label{eq:mom_con_supp} \\
    & \textstyle \sum_{n} \mathbb{L}[y_{n}^0](0) = 1,  \label{eq:mom_con_prob}\\
    & C_{\text{sym}} \textstyle \sum_{(n, i) \in \vs} u_{n^i} \mathbb{L}[y_{n, i}](s U_{q^b}(c)) \in h^b \\
    & \text{If FW :} \ \textstyle \sum_{(n, i) \in \vs} u_{n^i} \mathbb{L}[y_{n, i}](s U_{q^a}(c)) \in h^a \nonumber \\
    & \forall n \in 1..N, \  \gamma \in \N^{\circ 4}_{\leq 2\beta}, \ w(x) = x^\gamma: \\
    &\quad \mathbb{L}[y_n^0](w_0) + \mathbb{L}[y_{n,0}](\Lie_{n, 0}w) = \mathbb{L}[y^+_{n+1, 0}+y^-_{n-1, 0}](w) \nonumber\\   
    &\quad  \forall i \in 1..k/C_{\text{sym}}-1:\label{eq:mom_con_conservation}  \\
    & \quad \quad  
    \mathbb{L}[y^+_{n+1,i} + y^-_{n-1, i}](w_\phi)
    +   \mathbb{L}[y_{n,i}](\Lie_{n, i}w) \\
    & \quad \qquad \quad = \mathbb{L}[y^-_{n+1,i+1} + y^+_{n+1, i+1}](w)\nonumber\\
    &\quad \mathbb{L}[y^-_{n+1, k'} + y^+_{n-1, k'}](w_\phi) \\
    & \qquad + \mathbb{L}[\Lie_{n, d} w](y_{n, k'}) = \mathbb{L}[y^T_n](w_{2\pi/C_{\text{sym}}}).\nonumber
    \\
    &\ \forall \delta \in \N^{\circ 2}_{\leq 2\beta}: \quad \textstyle \sum_{(n, i) \in \vs} \mathbb{L}[y_{n, i}](c^{\gamma_1} s^{\gamma_2})  \label{eq:mom_con_uniformity}\\
    & \qquad = \textstyle \int_{\theta=0}^{2\pi/C_{\text{sym}}}  \cos(\theta)^{\delta_1} \sin(\theta)^{\delta_2} \ d\theta. \nonumber
\end{align}
\end{subequations}
\end{prob}
Constraint \eqref{eq:mom_con_prob} is a localizing/moment matrix constraint.
In order, terms \eqref{eq:mom_con_prob}-\eqref{eq:mom_con_uniformity} are pseudomoment-formulated versions of the Probability, Harmonics, Continuity, and Uniformity constraints (quarter-matching is omitted for simplicity of presentation). Because all support sets satisfy a ball constraint (as noted in Appendix \ref{app:mom_trunc}) and all pseudomoment variables are bounded, it holds that $\lim_{\beta \rightarrow \infty} = p^*$ \cite{putinar1993compact}.

The accuracy of the approximation $p^*_\beta$ to $p^*$ can be tightened by increasing $\beta$, and also by partitioning the state space $X$ into subregions according to the procedure in \cite{cibulka2021spatio} (e.g. splitting $B$ into $B([0, \pi]) \cup B([\pi, 2\pi])$).

%% file: sections/biography.tex
\begin{IEEEbiography}[{\includegraphics[width=1in,height=1.25in,clip,keepaspectratio]{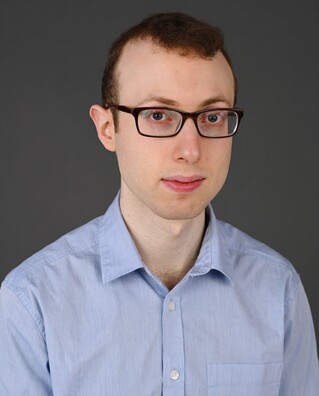}}]{Jared Miller} is currently a Postdoctoral Researcher at the Chair of Mathematical Systems Theory at the University of Stuttgart working with Prof. Carsten Scherer. He received his B.S. and M.S. degrees in Electrical Engineering from Northeastern University in 2018, and his Ph.D. degree from Northeastern University in 2023 under the advisorship of Mario Sznaier (Robust Systems Laboratory). He was previously a Postdoctoral Researcher Automatic Control Laboratory (IfA) at ETH Z\"{u}rich, in the research group of Prof. Roy S. Smith. He is a recipient of the 2020 Chateaubriand Fellowship from the Office for Science Technology of the Embassy of France in the United States. He was given an Outstanding Student Paper award at the IEEE Conference on Decision and Control in 2021 and in 2022. His research interests include renewable energy systems,  verification of nonlinear systems, and convex optimization.
\end{IEEEbiography}

\begin{IEEEbiography}[{\includegraphics[width=1in,height=1.25in,clip,keepaspectratio]{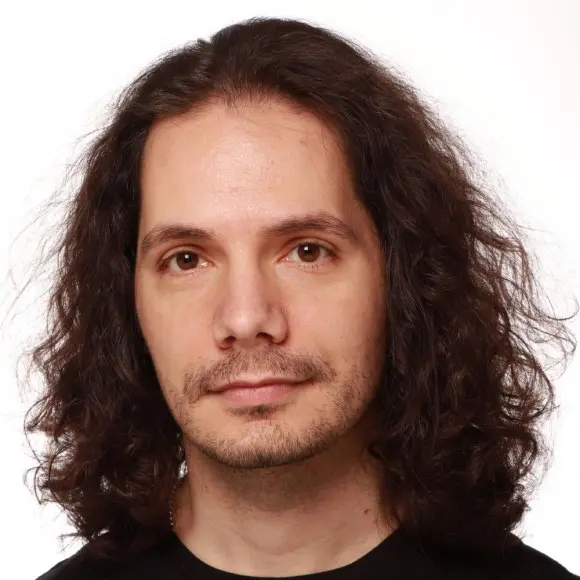}}]{Petros Karamanakos} received
the Diploma and Ph.D. degrees in electrical and
computer engineering from the National Technical
University of Athens (NTUA), Athens, Greece, in
2007, and 2013, respectively.
From 2010 to 2011 he was with the ABB Corporate Research Center, Baden-D¨attwil, Switzerland,
where he worked on model predictive control strategies for medium-voltage drives. From 2013 to 2016
he was a PostDoc Research Associate in the Chair
of Electrical Drive Systems and Power Electronics,
Technische Universit¨at M ¨unchen, Munich, Germany. Since 2016, he has been
with the Faculty of Information Technology and Communication Sciences,
Tampere University, Tampere, Finland, where he is currently an Associate
Professor. His main research interests lie at the intersection of optimal control,
mathematical programming, and power electronics, including model predictive
control and optimal modulation for utility-scale power converters and ac
variable speed drives.
Dr. Karamanakos has received three IEEE journal paper awards and five
prize paper awards at IEEE conferences. He serves as an Associate Editor
of the IEEE Transactions on Power Electronics, IEEE Journal of Emerging
and Selected Topics in Power Electronics, and IEEE Transactions on Industry
Applications. He has been a Regional Distinguished Lecturer of the IEEE
Power Electronics Society since 2022
\end{IEEEbiography}

\begin{IEEEbiography}[{\includegraphics[width=1in,height=1.25in,clip,keepaspectratio]{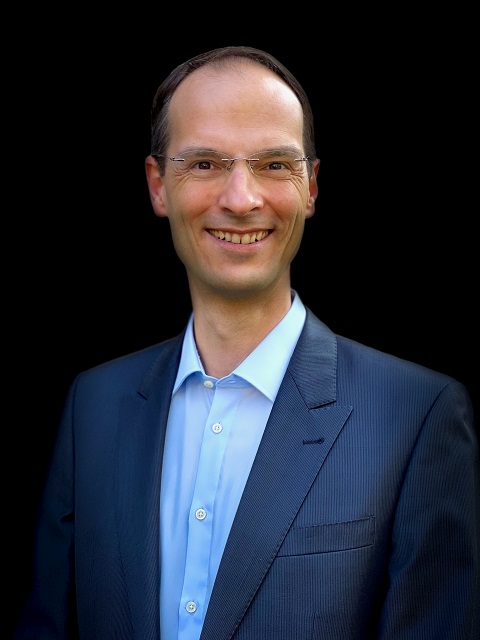}}]{Tobias Geyer} received the
Dipl.-Ing. degree in electrical engineering, the Ph.D.
in control engineering, and the Habilitation degree
in power electronics from ETH Zurich in the years
2000, 2005, and 2017, respectively.
After his Ph.D., he spent three years at GE Global
Research, Munich, Germany, three years at the
University of Auckland, Auckland, New Zealand,
and eight years at ABB’s Corporate Research Centre, Baden-D¨attwil, Switzerland. In 2020, he joined
ABB’s Medium-Voltage Drive division as R\&D platform manager of the ACS6080. In 2022, he became a Corporate Executive
Engineer. He has been an extraordinary Professor at Stellenbosch University,
Stellenbosch, South Africa, since 2017.
He is the author of more than 40 patent families, 160 publications and
the book “Model predictive control of high-power converters and industrial
drives” (Wiley, 2016). He teaches a regular course on model predictive control
at ETH Zurich. His research interests include medium-voltage and low-voltage
drives, utility-scale power converters, optimized pulse patterns and model
predictive control.
Dr. Geyer received the IEEE PELS Modeling and Control Technical
Achievement Award in 2022, the Semikron Innovation Award in 2021, and
the Nagamori Award in 2021. He also received two Prize Paper Awards of
IEEE transactions and three Prize Paper Awards at IEEE conferences. He is
a former Associate Editor of the IEEE Transactions on Industry Applications
(from 2011 until 2014) and the IEEE Transactions on Power Electronics (from
2013 until 2019). He was an International Program Committee Vice Chair of
the IFAC conference on Nonlinear Model Predictive Control in Madison, WI,
USA, in 2018. Dr. Geyer is a Distinguished Lecturer of the IEEE Power
Electronics Society from the year 2020 until 2023.

\end{IEEEbiography}

%% file: references.bib
@inproceedings{wachter2021convex,
  title={A convex relaxation approach for the optimized pulse pattern problem},
  author={Wachter, Lukas and Karaca, Orcun and Darivianakis, Georgios and Charalambous, Themistoklis},
  booktitle={2021 European Control Conference (ECC)},
  pages={2213--2218},
  year={2021},
  organization={IEEE}
}

@inproceedings{miller2025oppcdc,
      title={{Optimal Pulse Patterns through a Hybrid Optimal Control Perspective}}, 
      author={Jared Miller and Petros Karamanakos},
      year={2025},
      booktitle={2025 IEEE Conference on Decision and Control (CDC)},
      note={arxiv:2512.07585}, 
}

@book{barvinok2002convex,
 author = {Alexander Barvinok},
 title = {A Course in Convexity},
 publisher = {American Mathematical Society},
 year = {2002}
}

@article{miller2023peakhy,
  title={Peak estimation of hybrid systems with convex optimization},
  author={Miller, Jared and Sznaier, Mario},
  journal={Automatica},
  volume={183},
  pages={112641},
  year={2026},
  publisher={Elsevier}
}

@ARTICLE{zhao2017optimal,
  author={Zhao, Pengcheng and Mohan, Shankar and Vasudevan, Ram},
  journal={IEEE Transactions on Automatic Control}, 
  title={{Optimal Control of Polynomial Hybrid Systems via Convex Relaxations}}, 
  year={2020},
  volume={65},
  number={5},
  pages={2062-2077},
  doi={10.1109/TAC.2019.2929110}}

@article{lewis1980relaxation,
  title={{Relaxation of Optimal Control Problems to Equivalent Convex  Programs}},
  author={Lewis, RM and Vinter, RB},
  journal={Journal of Mathematical Analysis and Applications},
  volume={74},
  number={2},
  pages={475--493},
  year={1980},
  publisher={Elsevier}
}

@article{abu2025optimized,
  title={{Optimized Pulse Pattern Design for Electrical Drives by Deep Reinforcement Learning}},
  author={Abu-Ali, Mohammad and Berkel, Felix and Manderla, Maximilian and Pietschner, Markus and G{\"o}rges, Daniel},
  journal={Authorea Preprints},
  year={2025},
  publisher={Authorea}
}

@book{henrion2020moment,
  title={{The Moment-SOS Hierarchy
Lectures in Probability, Statistics, Computational Geometry, Control and Nonlinear PDEs}},
  author={Henrion, Didier and Korda, Milan and Lasserre, Jean Bernard},
  volume={4},
  year={2020},
  publisher={World Scientific}
}

@inproceedings{qashqai2020new,
  title={A new model-free space vector modulation technique for multilevel inverters based on deep reinforcement learning},
  author={Qashqai, Pouria and Al-Haddad, Kamal and Zgheib, Rawad},
  booktitle={Proc. {IEEE} Ind. Electron. Conf.},
  pages={2407--2411},
  address = {Singapore},
  year={2020},
  month = {Oct.},
}

@article{abu2025diff,
  title={{Optimized Pulse Pattern Design Using Differentiable Programming}},
  author={Abu-Ali, Mohammad and Strack, Moritz and Achterhold, Jan and Berkel, Felix and Manderla, Maximilian and Niepert, Mathias and G{\"o}rges, Daniel},
  journal={Authorea Preprints},
  year={2025},
  publisher={Authorea}
}

@article{rubio1975generalized,
  title={{Generalized Curves and Extremal Points}},
  author={Rubio, JE},
  journal={SIAM Journal on Control},
  volume={13},
  number={1},
  pages={28--47},
  year={1975},
  publisher={SIAM}
}

@phdthesis{tacchi2021thesis,
  title={{Moment-SOS hierarchy for large scale set approximation. Application to power systems transient stability analysis}},
  author={Tacchi, Matteo},
  year={2021},
  school={Toulouse, INSA}
}

@book{fattorini1999infinite,
  title={{Infinite Dimensional Optimization and Control Theory}},
  author={Fattorini, Hector O},
  volume={54},
  year={1999},
  publisher={Cambridge University Press}
}

@book{lasserre2009moments,
  title={{Moments, Positive Polynomials And Their} {Applications}},
  author={Lasserre, Jean Bernard.},
  isbn={9781908978271},
  series={Imperial College Press Optimization Series},
  year={2009},
  publisher={World Scientific Publishing Company}
}

@article{claeys2016modal,
  title={Modal occupation measures and {LMI} relaxations for nonlinear switched systems control},
  author={Claeys, Mathieu and Daafouz, Jamal and Henrion, Didier},
  journal={Automatica},
  volume={64},
  pages={143--154},
  year={2016},
  publisher={Elsevier}
}

@incollection{parrilo2005exploiting,
  title={{Exploiting Algebraic Structure in Sum of Squares Programs}},
  author={Parrilo, Pablo A},
  booktitle={Positive Polynomials in Control},
  pages={181--194},
  year={2005},
  publisher={Springer}
}

@article{putinar1993compact,
 ISSN = {00222518, 19435258},
 author = {Mihai Putinar},
 journal = {Indiana University Mathematics Journal},
 number = {3},
 pages = {969--984},
 publisher = {Indiana University Mathematics Department},
 title = {{Positive Polynomials on Compact Semi-algebraic Sets}},
 volume = {42},
 year = {1993}
}

@article{zhang2001zeno,
  title={Zeno hybrid systems},
  author={Zhang, Jun and Johansson, Karl Henrik and Lygeros, John and Sastry, Shankar},
  journal={Int. J. Robust Nonlinear Control},
  volume={11},
  number={5},
  pages={435--451},
  year={2001},
  publisher={Wiley Online Library}
}

@misc{teel2012hybrid,
  title={{Hybrid Dynamical Systems: Modeling, Stability, and Robustness}},
  author={Teel, Andrew R},
  year={2012},
  publisher={Princeton University Press}
}

@article{holtz2002pulsewidth,
  title={Pulsewidth modulation---{A} survey},
  author={Holtz, Joachim},
  journal={IEEE Trans. Ind. Electron.},
  volume={39},
  number={5},
  pages={410--420},
  year={2002},
  month = {Dec.},
  publisher={IEEE}
}

@article{buja1980optimum,
  title={Optimum Output Waveforms in {PWM} Inverters},
  author={Buja, Giuseppe S},
  journal={IEEE Trans. Ind. Appl.},
  volume = {IA-16},
  number={6},
  pages={830--836},
  year={1980},
  month={Nov./Dec.}
}

@book{puterman1990markov,
  title={Markov decision processes: discrete stochastic dynamic programming},
  author={Puterman, Martin L},
  year={2014},
  publisher={John Wiley \& Sons}
}

@article{patel1974generalized,
  title={Generalized Techniques of Harmonic Elimination and Voltage Control in Thyristor Inverters: {Part II}---{Voltage} Control Techniques},
  author={Patel, Hasmukh S and Hoft, Richard G},
  journal={IEEE Trans. Ind. Appl.},
  volume = {IA-10},
  number={5},
  pages={666--673},
  year={1974},
  month={Sep./Oct.},
}

@ARTICLE{kavouski2012bee,
  author={Kavousi, Ayoub and Vahidi, Behrooz and Salehi, Reza and Bakhshizadeh, Mohammad Kazem and Farokhnia, Naeem and Fathi, S. Hamid},
  journal={IEEE Trans. Power Electron.}, 
  title={Application of the Bee Algorithm for Selective Harmonic Elimination Strategy in Multilevel Inverters}, 
  year={2012},
  volume={27},
  number={4},
  pages={1689--1696},
  month = {Apr.},
  doi={10.1109/TPEL.2011.2166124}}

@article{patel1973generalized,
  title={Generalized Techniques of Harmonic Elimination and Voltage Control in Thyristor Inverters: {Part I}---{Harmonic} Elimination},
  author={Patel, Hasmukh S and Hoft, Richard G},
  journal={IEEE Trans. Ind. Appl.},
  volume = {IA-9},
  number={3},
  pages={310--317},
  year={1973},
  month = {May/Jun.}
}

@article{chiasson2004unified,
  title={A unified approach to solving the harmonic elimination equations in multilevel converters},
  author={Chiasson, John N and Tolbert, Leon M and McKenzie, Keith J and Du, Zhong},
  journal={IEEE Trans. Power Electron.},
  volume={19},
  number={2},
  pages={478--490},
  year={2004},
  month = {Mar.},
  publisher={IEEE}
}

@article{wells2005selective,
  title={Selective harmonic control: {A} general problem formulation and selected solutions},
  author={Wells, Jason R and Nee, Brett M and Chapman, Patrick L and Krein, Philip T},
  journal={IEEE Trans. Power Electron.},
  volume={20},
  number={6},
  pages={1337--1345},
  year={2005},
  month = {Nov.},
  publisher={IEEE}
}

@article{dahidah2014review,
  title={A Review of Multilevel Selective Harmonic Elimination {PWM}: Formulations, Solving Algorithms, Implementation and Applications},
  author={Dahidah, Mohamed SA and Konstantinou, Georgios and Agelidis, Vassilios G},
  journal={IEEE Trans. Power Electron.},
  volume={30},
  number={8},
  pages={4091--4106},
  year={2015},
  month = {Aug.},
  publisher={IEEE}
}

@ARTICLE{Jahns2001drives,
  author={Jahns, T. M. and Blasko, V.},
  journal={Proc. IEEE}, 
  title={Recent advances in power electronics technology for industrial and traction machine drives}, 
  year={2001},
  volume={89},
  number={6},
  pages={963--975},
  month = {Jun.},
}

@inproceedings{meili2006optimized,
  title={Optimized Pulse Patterns for the 5-Level {ANPC} Converter for High Speed High Power Applications},
  author={Meili, Jorg and Ponnaluri, Srinivas and Serpa, Leonardo and Steimer, Peter K and Kolar, Johann W},
  booktitle={Proc. {IEEE} Ind. Electron. Conf.},
  pages={2587--2592},
  year={2006},
  month = {Nov.},
  address = {Paris, France},
}

@article{birth2019generalized,
  title={Generalized Three-Level Optimal Pulse Patterns With Lower Harmonic Distortion},
  author={Birth, Annika and Geyer, Tobias and du Toit Mouton, Hendrik and Dorfling, Martinus},
  journal={IEEE Trans. Power Electron.},
  volume={35},
  number={6},
  pages={5741--5752},
  year={2019},
  month = {Jun.},
  publisher={IEEE}
}

@ARTICLE{dorfling2024jucntion,
  author={Dorfling, Tinus and Geyer, Tobias},
  journal={IEEE Trans. Power Electron.}, 
  title={Thermally Constrained Optimized Pulse Patterns for Medium-Voltage Neutral-Point-Clamped Converters}, 
  year={2024},
  volume={39},
  number={10},
  pages={13160--13176},
  month = {Oct.},
}

@inproceedings{abate2021fossil,
  title={{FOSSIL}: a software tool for the formal synthesis of lyapunov functions and barrier certificates using neural networks},
  author={Abate, Alessandro and Ahmed, Daniele and Edwards, Alec and Giacobbe, Mirco and Peruffo, Andrea},
  booktitle={Proceedings of the 24th international conference on hybrid systems: computation and control},
  pages={1--11},
  year={2021}
}

@inproceedings{gao2013dreal,
  title={{dReal: An SMT Solver for Nonlinear Theories over the Reals}},
  author={Gao, Sicun and Kong, Soonho and Clarke, Edmund M},
  booktitle={International Conference on Automated Deduction},
  pages={208--214},
  year={2013},
  organization={Springer}
}

@article{mohajerin2018infinite,
  title={{From Infinite to Finite Programs: Explicit Error Bounds with Applications to Approximate Dynamic Programming}},
  author={Mohajerin Esfahani, Peyman and Sutter, Tobias and Kuhn, Daniel and Lygeros, John},
  journal={SIAM J. Optim.},
  volume={28},
  number={3},
  pages={1968--1998},
  year={2018},
  publisher={SIAM}
}

@manual{mosek110,
   author = "MOSEK ApS",
   title = "The MOSEK optimization toolbox for MATLAB manual. Version 11.0.6.",
   year = 2025
 }

@INPROCEEDINGS{lofberg2004yalmip,
  author={J. {Lofberg}},
  booktitle={ICRA (IEEE Cat. No.04CH37508)}, 
  title={{YALMIP}: a toolbox for modeling and optimization in {MATLAB}}, 
  year={2004},
  volume={},
  number={},
  pages={284-289},}

@article{henrion2009gloptipoly,
  title={Glopti{P}oly 3: moments, optimization and semidefinite programming},
  author={Henrion, Didier and Lasserre, Jean-Bernard and L{\"o}fberg, Johan},
  journal={Optimization Methods \& Software},
  volume={24},
  number={4-5},
  pages={761--779},
  year={2009},
  publisher={Taylor \& Francis}
}

@book{zhong2012control,
  title={Control of Power Inverters in Renewable Energy and Smart Grid Integration},
  author={Zhong, Qing-Chang and Hornik, Tomas},
  year={2012},
  publisher={John Wiley \& Sons}
}

@article{dahidah2008hybrid,
  title={Hybrid genetic algorithm approach for selective harmonic control},
  author={Dahidah, Mohamed SA and Agelidis, Vassilios G and Rao, Machavaram V},
  journal={Energy Convers. and Manag.},
  volume={49},
  number={2},
  pages={131--142},
  year={2008},
  month = {Feb.},
  publisher={Elsevier}
}

@article{wang2022application,
  title={{Application of Newton Identities in Solving Selective Harmonic Elimination Problem With Algebraic Algorithms}},
  author={Wang, Chenxu and Zhang, Qi and Chen, Dunzhi and Li, Zhaoyuan and Yu, Wensheng and Yang, Kehu},
  journal={IEEE Journal of Emerging and Selected Topics in Power Electronics},
  volume={10},
  number={5},
  pages={5870--5881},
  year={2022},
  publisher={IEEE}
}

@inproceedings{wang2021sparsejsr,
  title={{SparseJSR: A fast algorithm to compute joint spectral radius via sparse SOS decompositions}},
  author={Wang, Jie and Maggio, Martina and Magron, Victor},
  booktitle={2021 American Control Conference (ACC)},
  pages={2254--2259},
  year={2021},
  organization={IEEE}
}

@article{parrilo2008approximation,
  title={Approximation of the joint spectral radius using sum of squares},
  author={Parrilo, Pablo A and Jadbabaie, Ali},
  journal={Linear Algebra and its Applications},
  volume={428},
  number={10},
  pages={2385--2402},
  year={2008},
  publisher={Elsevier}
}

@article{manchester2011regions,
  title={{Regions of Attraction for Hybrid Limit Cycles of Walking Robots}},
  author={Manchester, Ian R and Tobenkin, Mark M and Levashov, Michael and Tedrake, Russ},
  journal={IFAC Proceedings Volumes},
  volume={44},
  number={1},
  pages={5801--5806},
  year={2011},
  publisher={Elsevier}
}

@inproceedings{prajna2004safety,
  title={{Safety Verification of Hybrid Systems Using Barrier Certificates}},
  author={Prajna, Stephen and Jadbabaie, Ali},
  booktitle={International workshop on hybrid systems: Computation and control},
  pages={477--492},
  year={2004},
  organization={Springer}
}

@article{ahmadi2019polynomial,
  title={Polynomial {N}orms},
  author={Ahmadi, Amir Ali and De Klerk, Etienne and Hall, Georgina},
  journal={SIAM Journal on Optimization},
  volume={29},
  number={1},
  pages={399--422},
  year={2019},
  publisher={SIAM}
}

@ARTICLE{miller2024chancepeak,
  author={Miller, Jared and Tacchi, Matteo and Sznaier, Mario and Jasour, Ashkan},
  journal={IEEE Transactions on Automatic Control}, 
  title={{Convex Computation of Value-at-Risk Bounds for Stochastic Processes}}, 
  year={2025},
    volume={70},
  number={6},
  pages={3936-3951},
  doi={10.1109/TAC.2024.3519977}}

@book{sommese2005numerical,
  title={{The Numerical Solution of Systems of Polynomials Arising in Engineering and Science}},
  author={Sommese, Andrew J and Wampler, Charles W.},
  year={2005},
  publisher={World Scientific}
}

@article{yang2016unified,
  title={{Unified Selective Harmonic Elimination for Multilevel Converters}},
  author={Yang, Kehu and Zhang, Qi and Zhang, Jianjun and Yuan, Ruyi and Guan, Qiang and Yu, Wensheng and Wang, Jin},
  journal={IEEE Transactions on Power Electronics},
  volume={32},
  number={2},
  pages={1579--1590},
  year={2016},
  publisher={IEEE}
}

@article{ali2024optimal,
  title={{Optimal Control of Switched Dynamical Systems Under Dwell Time Constraints-Theory and Computation}},
  author={Ali, Usman and Egerstedt, Magnus},
  journal={Trans. Autom. Control},
  volume={70},
  number={4},
  pages={2362--2373},
  year={2025},
  month = {Apr.},
  publisher={IEEE}
}

@inproceedings{koukoula2024fast,
  title={Fast Computation of Optimized Pulse Patterns for Multilevel Converters},
  author={Koukoula, Isavella and Karamanakos, Petros and Geyer, Tobias},
  booktitle={Proc. {IEEE} Energy Convers. Congr. Expo.},
  pages={4352--4354},
  year={2024},
  month = {Oct.},
  address = {Phoenix, AZ, USA},
  organization={IEEE}
}

@book{holmes2003pulse,
  title={Pulse Width Modulation for Power Converters: Principles and Practice},
  author={Holmes, D Grahame and Lipo, Thomas A},
  year={2003},
  publisher={John Wiley \& Sons}
}

@inproceedings{breiding2018homotopycontinuation,
  title={{HomotopyContinuation. jl: A package for homotopy continuation in Julia}},
  author={Breiding, Paul and Timme, Sascha},
  booktitle={International congress on mathematical software},
  pages={458--465},
  year={2018},
  organization={Springer}
}

@article{eago2022,
author = {Wilhelm, M.E. and Stuber, M.D.},
title = {{EAGO.jl: Easy Advanced Global Optimization in Julia}},
journal = {Optimization Methods and Software},
volume = {37},
number = {2},
pages = {425-450},
year  = {2022},
publisher = {Taylor & Francis}
}

@article{toubal2022neural,
  title={A neural network-based selective harmonic elimination scheme for five-level inverter},
  author={Toubal Maamar, Alla Eddine and Helaimi, M'hamed and Taleb, Rachid and Kermadi, Mostefa and Mekhilef, Saad},
  journal={Int. J. Circ. Theory and Appl.},
  volume={50},
  number={1},
  pages={298--316},
  year={2022},
  publisher={Wiley Online Library}
}

@article{cibulka2021spatio,
  title={{Spatio-Temporal Decomposition of Sum-of-Squares Programs for the Region of Attraction and Reachability}},
  author={Cibulka, Vit and Korda, Milan and Hani{\v{s}}, Tom{\'a}{\v{s}}},
  journal={IEEE Control Systems Letters},
  volume={6},
  pages={812--817},
  year={2021},
  publisher={IEEE}
}

@article{kouro2010recent,
  title={Recent Advances and Industrial Applications of Multilevel Converters},
  author={Kouro, Samir and Malinowski, Mariusz and Gopakumar, K and Pou, Josep and Franquelo, Leopoldo G and Wu, Bin and Rodriguez, Jose and P{\'e}rez, Marcelo A and Leon, Jose I},
  journal={IEEE Trans. Ind. Electron.},
  volume={57},
  number={8},
  pages={2553--2580},
  year={2010},
  month = {Aug.},
  publisher={IEEE}
}

@article{koukoula2024optimal,
  title={Optimal pulse width modulation of three-level converters with reduced common-mode voltage},
  author={Koukoula, Isavella and Karamanakos, Petros and Geyer, Tobias},
  journal={IEEE Trans. Ind. Appl.},
  volume={60},
  number={3},
  pages={4062--4075},
  year={2024},
  month = {May/Jun.},
  publisher={IEEE}
}

@article{koukoula2025losses,
  author={Koukoula, Isavella and Karamanakos, Petros and Geyer, Tobias},
  journal={IEEE Trans. Ind. Appl.}, 
  title={Loss-Constrained Three-Level Optimized Pulse Patterns With Robustness to Power Factor Variations}, 
  year={2025},
  volume={61},
  number={4},
  pages={6511--6523},
  month = {Jul./Aug.},
}

@ARTICLE{koukoula2025junction,
  author={Koukoula, Isavella and Karamanakos, Petros and Geyer, Tobias},
  journal={IEEE Trans. Power Electron.}, 
  title={Three-Level Optimized Pulse Patterns With Bounded Junction Temperature and Relaxed Properties}, 
  year={2025, in press, DOI: 10.1109/TPEL.2025.3626863},
  volume={},
  number={},
  pages={1--13},
}

@article{geyer2023optimized,
  title={Optimized Pulse Patterns With Bounded Semiconductor Losses},
  author={Geyer, Tobias and Karamanakos, Petros and Koukoula, Isavella},
  journal={IEEE Trans. Power Electron.},
  volume={39},
  number={3},
  pages={3233--3243},
  year={2023},
  month = {Mar.},
  publisher={IEEE}
}

@article{rahmanpour2023three,
  title={Harmonic-Constrained Three-Level Optimized Pulse Patterns for Grid-Connected Converters With ${LCL}$ Filters}, 
  author={Rahmanpour, Shirin and Karamanakos, Petros and Geyer, Tobias},
  journal={IEEE Trans. Ind. Appl.}, 
  year={2025},
  volume={61},
  number={5},
  pages={7481-7492},
  month={Sep./Oct.},
}
